\documentclass[ams,onecolumn]{emulateapj}

\usepackage[OT2,OT1]{fontenc}
\newcommand\cyrtext[1]{{\fontencoding{OT2}\selectfont #1}}
\usepackage{amsmath}
\usepackage{amsthm}
\usepackage{amssymb}
\usepackage{amsfonts}
\newenvironment{demo}{{\noindent\small\bf{\underline{Proof}:}}}{\begin{flushright}$\blacksquare$
\end{flushright}}
\newcommand{\e}{\mathop{\rm e}\nolimits}
\newtheorem{thm}{Theorem}[section]
\newtheorem{prop}[thm]{Proposition}

\shorttitle{The CHEF basis: a new tool for image analysis}
\shortauthors{Jim\'enez-Teja \& Ben\'\i tez}

\begin{document}

\title{A new tool for image analysis based on Chebyshev rational functions: CHEF functions}

\author{Y. Jim\'enez-Teja and N. Ben\'\i tez}

\affil{Instituto de Astrof\'\i sica de Andaluc\'\i a (CSIC), Glorieta de la Astronom\'\i a s/n, Granada, E-18008, Spain}
\email{yojite@iaa.es}

\begin{abstract}

  We introduce a new approach to the modelling of the light distribution of galaxies, an orthonormal polar base formed by a combination of Chebyshev rational functions and Fourier polynomials that we call CHEF functions, or CHEFs. We have developed an orthonormalization process to apply this basis to pixelized images, and implemented the method as a Python pipeline. 
   
  The new basis displays remarkable flexibility, being able to accurately fit all kinds of galaxy shapes, including irregulars, spirals, ellipticals, highly compact and highly elongated galaxies. It does this while using fewer components that similar methods, as shapelets, and without producing artifacts, due to the efficiency of the rational Chebyshev polynomials to fit quickly decaying functions like galaxy profiles. The method is lineal and very stable, and therefore capable of processing large numbers of galaxies in a fast and automated way. 

 Due to the high quality of the fits in the central parts of the galaxies, and the efficiency of the CHEF basis modeling galaxy profiles up to very large distances, the method provides highly accurate estimates of total galaxy fluxes and ellipticities. Future papers will explore in more detail the application of the method to perform multiband photometry, morphological classification and weak shear measurements.  

\end{abstract}


\section{Introduction}

 Large surveys play an increasingly central role in Astronomy. The Sloan Digital Sky Survey (SDSS) \citep{sdss} pioneered this trend, obtaining multi-color images covering more than a quarter of the sky, with a final dataset of 230M celestial objects in an area of 8,400 square degrees, during its eight years of operations. The new generation of surveys is even more ambitious, and will generate astonishing data volumes. The Panoramic Survey Telescope \& Rapid Response System (Pan-STARRS) in Hawaii \citep{pan_starrs}, composed by four individual optical systems of 1.8 meter diameter each, will be able to survey 6,000 deg$^2$ per night. It will catalog 99\% of the stars in the northern hemisphere and make an ultradeep survey of 1200 square degrees to $g=27$. The Dark Energy Survey (DES) \citep{des} will study the acceleration of the expanding universe by surveying 5,000 deg$^2$ of the southern sky in five bands, and expects to detect more than 300 million galaxies. The Javalambre-PAU Astrophysical Survey (J-PAS) project \citep{jpas} will survey 8,000 deg$^2$ in 40 different bands, starting in 2013, measuring  photometric redshifts with $dz/(1+z)\sim 0.003$ and $dz/(1+z)\sim 0.01$ for respectively, 100M and 300M galaxies. The Large Synoptic Survey Telescope (LSST) \citep{lsst} will start to cover an area of 30,000 deg$^2$ in 2015 with six optical filters and expects to find a number of galaxies, 10 billion, dwarfing all other projects. To analyze and exploit such huge datasets, it is therefore necessary to develop techniques which are automated, objective and efficient. 

  Modeling of the light distribution of galaxies is a crucial step for a wide array of problems in Galaxy Evolution and Observational Cosmology, which require accurate galaxy shape and flux measurements. The use of analytical profiles is the approach with longest tradition, employed e.g. by GIM2D \citep{gim2d}. This IRAF package analyzes galaxy images with a bulge/disk decomposition consisting of a S\'ersic profile \citep{sersic} plus an exponential, and optimization of the parameters is performed by the Metropolis algorithm. Other works try to adjust galaxy profiles as a sum of elliptical Gaussians, as can be observed in \citet{ellipticals1} or in im2shape method \citep{ellipticals2}, and apply a maximum likelihood method to fit them. Lensing work  have also contributed to develop new techniques, as \citet{pol1} or \citet{pol2}, where a Gaussian times a polynomial profile is used as a model function. The idea of doing the model fitting in Fourier space has been also tried, for example in \citet{lensfit}, being based on a de Vaucouleurs profile. 

  The most widely used implementation of this approach is perhaps GALFIT \citep{peng,peng2}, which combines analytical profiles for the radial profile and trigonometric functions for the angular distribution. It works extremely well with objects with smooth features, as elliptical galaxies and thanks to its recently introduced angular components it displays considerable flexibility. However, as most analytical approaches,  it requires considerable interactivity from the user in order to model complex galaxy profiles, especially at high S/N, since the number, type and initial values of the components have to be adjusted manually. Due to the large number of degrees of freedom, and the non-linearity of the fitting functions, which may require long convergence times or may not converge at all, it will be difficult to implement GALFIT, or any other analytical profile-based approaches, as a high-precision tool to handle the hundreds of millions of galaxies coming from the future large imaging surveys. 

  A different approach, much better adapted to blind, automated fitting, is the use of orthonormal bases. The widely-known wavelets bases have been profusely applied to a great variety of science fields, including Astrophysics. They form a complete and orthonormal system across scales, capable of a multiscale decomposition of galaxies \citep{wavelets}. They have inspired  other bases, for example, beamlets \citep{beamlets} or curvelets \citep{curvelets}, used  to efficiently represent certain features present in astronomical images. 

  The most succesful orthonormal basis methods in Astronomy are the so called shapelets \citep{massey,refregier}. Built using Hermite polynomials and Gaussians, these bases are extremely flexible, capable of decomposing irregular objects and galaxies displaying small features. They are also fast and lineal. However, they feature an exponentially squared cut-off on the radial basis functions, which makes them fall significantly faster than most galaxy profiles. Because of this, they  require very large numbers of coefficients to model objects with extended wings, what leads to oscillations in the radial shape \citep{melchior}. This limitation seriously hampers their usefulness as automatic estimators of shapes and fluxes for large galaxy surveys, although they have been used for many astrophysical applications \citep{ellipticals1,refregier,massey2,kelly,kelly2,goldberg,heymans,massey3,kuijken2}. An interesting attempt to tackle this issue are sersiclets \citep{sersiclets}, a polar basis based on S\'ersic profiles. However, S\'ersic functions are poorly suited to form a 2D orthogonal basis, and this basis cannot fit irregular galaxy profiles. 

  Given the obvious advantages of orthonormal bases, it seems highly desirable to develop one with the morphological flexibility and processing speed of the shapelets, but which is able to accurately and compactly represent realistic galaxy profiles up to very large radii. Here we introduce such a basis, which we called CHEF functions or ``CHEFs'', built using Chebyshev polynomials \citep{boyd}, which, as it is well known, display remarkable properties of efficiency when fitting functions in the [-1,1] interval as their minimax property states \citep{mason}. The CHEF functions are separately built in polar coordinates, by expanding the radial component using Chebyshev rational functions (the composition of a Chebyshev polynomial and an algebraical mapping) and developing the azimuthal coordinate by means of sines and cosines. The result is a complete and orthonormal set of basis functions very well suited to fit astronomical objects able to fully model the whole flux of a galaxy to very large radii using a small number of coefficients. To explore the reliability and performance of this new technique, we have tested it on a wide sample of both simulated and real data, comparing the results with those obtained by the most popular publicly available techniques, GALFIT and IDL Shapelets software.

  The paper is organized as follows: in Sec. 2., we present the mathematical definition of the CHEF basis, demonstrate the fast decay rate of the coefficients, what ensures the compactness of the basis, and derive the main galaxy parameters from the decomposition coefficients. Sec.3. focuses on the practical implementation of the algorithm, describing how to discretize the basis and choose the free parameters of the fit, namely the scale size, the center and the decomposition order, and describes our functioning pipeline, which is being used to process data from different surveys. Sec. 4. shows the application of the method to different datasets, both real and simulated, and discusses its performance compared with that of shapelets. Sec. 5. states the main conclusions of the paper.  

\section{Mathematical definitions and results}
\label{math}
  We present an orthonormal basis set for image analysis based on polar basis functions that are constructed from rational
Chebyshev functions, defined on the semi-infinite interval $[0,+\infty)$, to expand the radial coordinate, and from sine and cosine waves to represent the azimuthal component. We call this basis ``CHEF'' functions (from \cyrtext{Ch}, 
the initial letter of Chebyshev, which is pronounced as ''Che'' in russian and F, the first letter of Fourier's family name) or, more compactly, ``CHEFs''. 

 The \textit{Chebyshev polynomial} of degree $n$ is
defined by the relation:
\begin{equation}
T_n(z)=\cos(n\,\theta) \label{Cheby},\hspace{0.5cm} \mbox{where }
z=\cos\,\theta.
\end{equation}

  Chebyshev polynomials constitute a basis of the
$L^2$-space of squared-integrable functions defined on the finite
interval $[-1,1]$. Using an algebraic coordinate transformation,
this domain can be mapped onto the semi-infinite interval
$\left[0,+\infty\right)$ and the \textit{rational Chebyshev
functions} can be obtained \citep{boyd}:
\begin{equation}\displaystyle
TL_n(r)=T_n\left(\frac{r-L}{r+L}\right)=\cos\left(n\cdot\arccos\left(
\frac{r-L}{r+L}\right)\right).\label{rational_Cheby}
\end{equation}

\begin{figure}[h]
\begin{center}
\includegraphics[width=9cm]{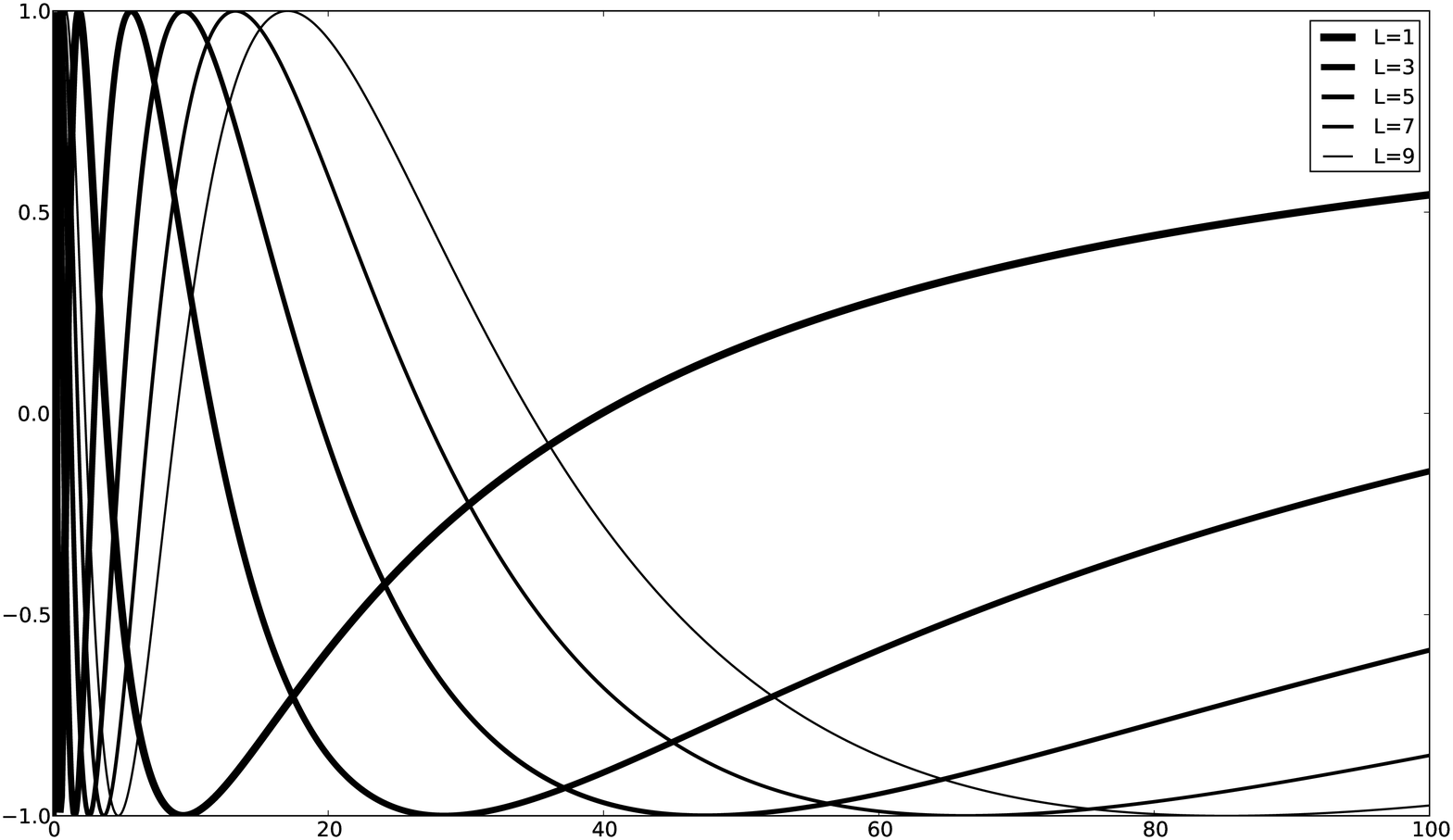}
\caption{Chebyshev rational function of order $n=5$ with different scale sizes.}\label{fig1}
\end{center}
\end{figure}

  The constant $L$ appearing in this expression will be called the \textit{scale parameter} from now on and it is related to the ``width'' of the functions, and it does not affect their shape, as it can be seen in figure \ref{fig1}. This scale is a free 
parameter of the decomposition and it has to be determined when applying the method to real data. Luckily, as we see below, the quality of the fit is not very sensitive to its exact value.

 These rational Chebyshev functions are orthogonal with respect to the weight function $\displaystyle
f(r)=\frac{1}{r+L}\sqrt{\frac{L}{r}}$, that is:
\begin{equation}
\int\limits_0^{\infty} TL_{i}(r;L)\,TL_{j}(r;L)\,\frac{1}{r+L}
\sqrt{\frac{L}{r}}\;dr=\left\{\begin{array}{ll} \pi, & \mbox{if }
i=0 \\ 0, & \mbox{if } i\neq j \\ \pi/2, & \mbox{if }
j>0
\end{array}\right..\label{orthogonality}
\end{equation}

\begin{figure}
\centerline{\includegraphics[width=10cm]{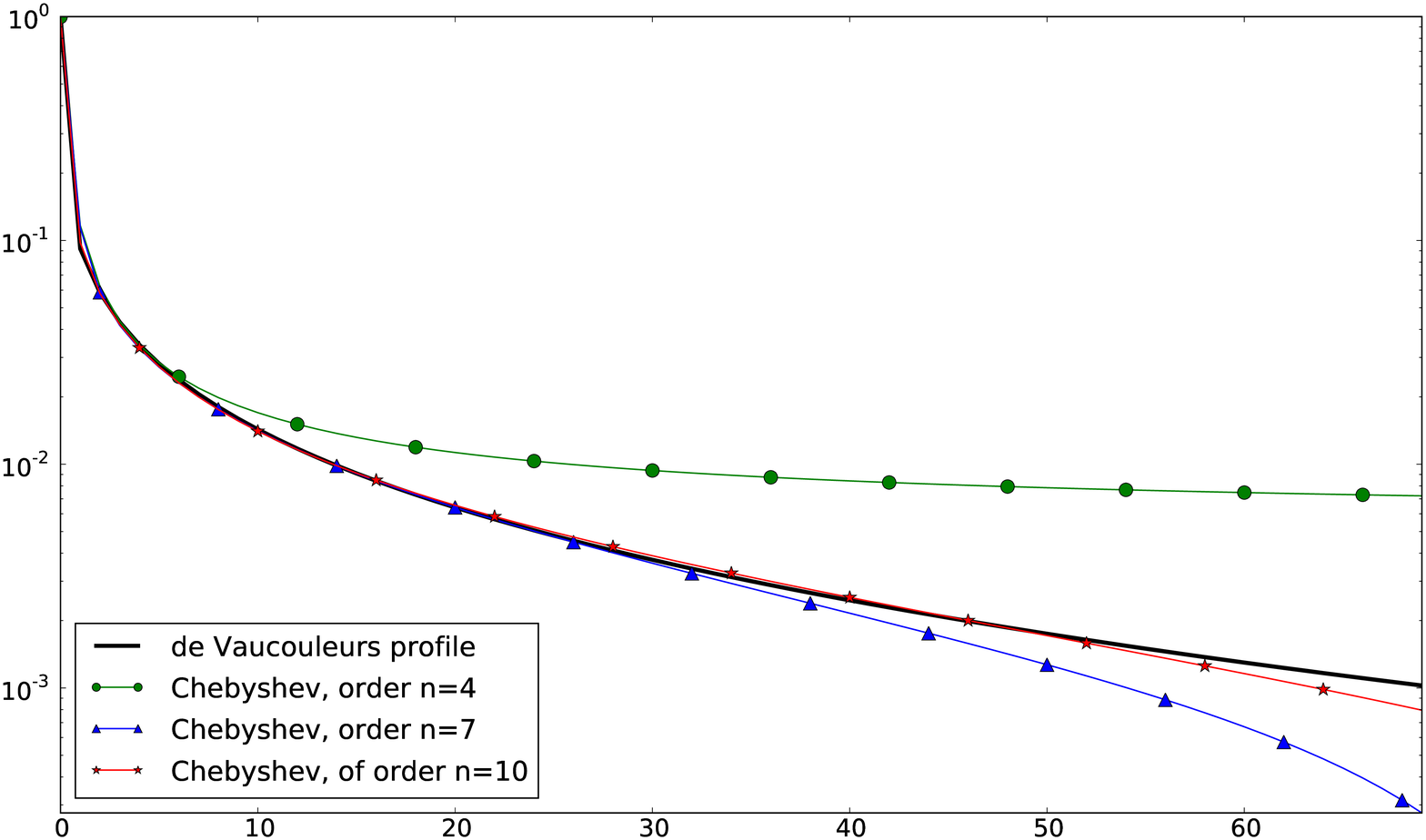}}
\centerline{\includegraphics[width=10cm]{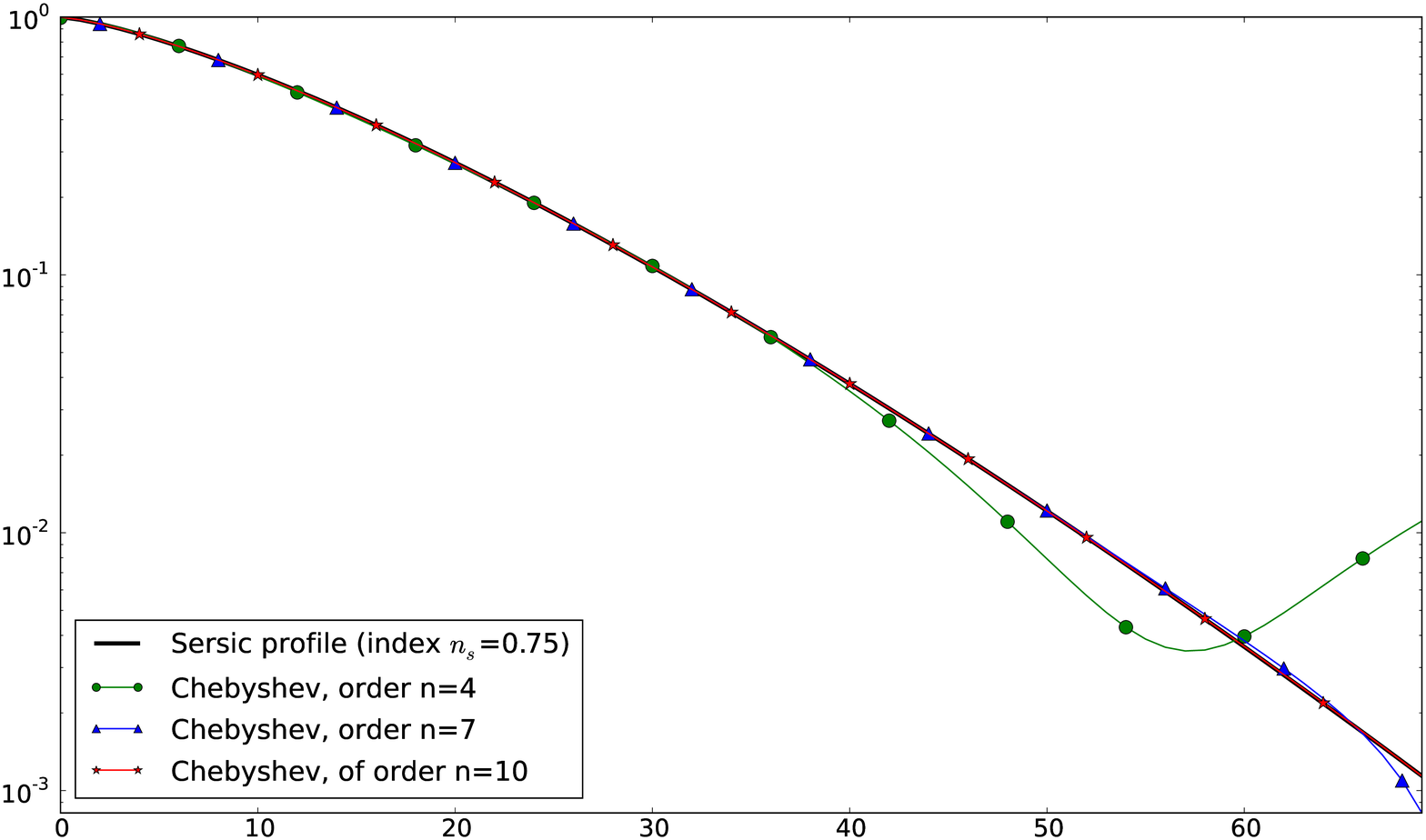}}
\caption{Chebyshev approximations to both a de Vaucouleurs (top panel) and a S\'ersic profile (lower panel) (with index $n_s=0.75$), using Chebyshev rational functions of different order $n$. Order $n\lesssim 10$ is enough to provide high accuracy over the scales sampled by typical astronomical images.}
\label{approx}
\end{figure}

 They tend monotously to 1 after their last minimum, more and more slowly as the order grows. Because of this property, and unlike shapelets, which fall off like a Gaussian, they can fit extended galaxy wings very compactly. At small radii, they quickly oscillate, and are thus capable of accurately describing the very fast change of flux with radius displayed by most galaxies. The frontier between both regimes is roughly set by the scale parameter $L$. Their efficiency in modeling galaxy profiles is illustrated by figure \ref{approx} which shows the rational Chebyshev approximations to both a de Vaucouleur and a S\'ersic profile (with index $n_s=0.75$), using different Chebyshev rational functions orders. It can be observed how they accurately match both profiles over 3 orders of magnitude with less than 10 coefficients. In comparison, shapelets fail to model a de Vaucouleurs profile over two orders of magnitude even going to order 16. (see fig. 2 from \citep{jim}). 

\setlength{\unitlength}{1cm}

  The polar CHEF functions are separable in $r$ and $\theta$, and built by expanding the
radial component with normalized rational Chebyshev functions, whereas the angular coordinate is represented by a combination of sines and cosines:
\begin{equation}\displaystyle
\left\{\phi_{nm}(r,\theta;L)\right\}_{nm} =\left\{\frac{C}{\pi}\;TL_{n}(r) 
W_{m}(\theta)\right\},\label{basis}
\end{equation}

\noindent where $\displaystyle C=\left\{
\begin{array}{ll} 1, & \mbox{if } n=0 \\ 2, & \mbox{if } n>0 \end{array}\right.$, and
$W_{m}(\theta)$ is a general
expression to represent both $\sin{(m\theta)}$ and $\cos{(m\theta)}$.\\
\setlength{\unitlength}{1cm}
\begin{figure}%[h]
\begin{center}
\includegraphics[width=9cm]{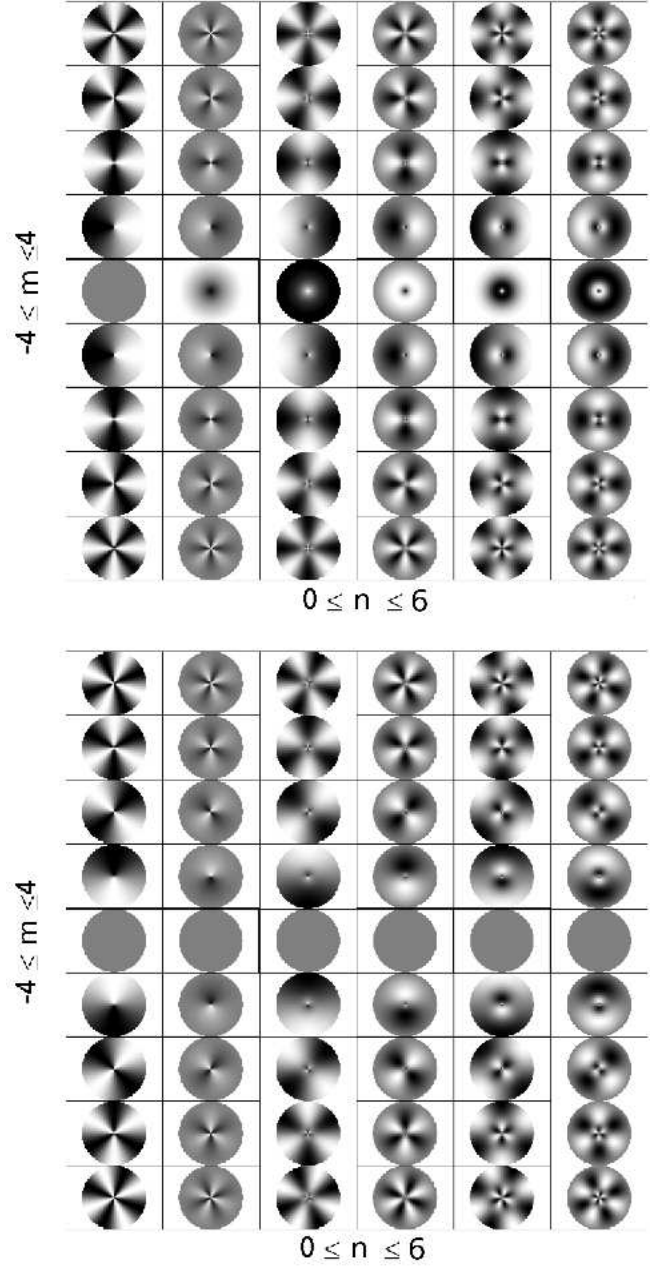}
\caption{Cosine and sine components of the first CHEF basis functions with scale size $L=1$.}
\end{center}\label{fig2}
\end{figure}

  CHEFs with scale size $L=1$ and
coefficient indices ranging from $0\leq n\leq 6$ and $-4\leq
m\leq 4$ are displayed in figure \ref{fig2}. This set forms an
orthonormal basis of the space of the finite-energy functions
${L}^2\left(\left[0,+\infty\right)\times
\left[-\pi,\pi\right],\langle\cdot,\cdot\rangle\right)$, which
constitutes a Hilbert space with the inner product defined by
\begin{equation}
\langle f,g\rangle=\int\limits_0^{+\infty}\int\limits_{-\pi}^\pi
f\left(r,\theta\right)\,\overline{g\left(r,\theta\right)}\,\frac{1}{r+L}
\sqrt{\frac{L}{r}}\,d\theta\,dr. \label{inner_product}
\end{equation}

  In this way, a smooth enough function $f$ in polar
coordinates can be decomposed into
\begin{equation}
f\left(r,\theta\right)= \displaystyle\frac{C}{2\pi^2}\sum\limits_{m=0}^{+\infty}
\sum\limits_{n=0}^{+\infty}f_{nm}\;
TL_{n}\left(r\right)W_{m}\left(\theta\right)=\displaystyle\frac{C}{2\pi^2}\sum\limits_{m=0}^{+\infty}
\sum\limits_{n=0}^{+\infty}\left[f_{nm}^c\;
TL_{n}\left(r\right)\cos{\left(m\theta\right)}+f_{nm}^s\;
TL_{n}\left(r\right)\sin{\left(m\theta\right)}\right]\label{descomp}
\end{equation}

\noindent and the coefficients $f_{nm}$ are calculated by
\begin{equation}\begin{array}{l}
\displaystyle f_{nm}^c=\frac{C}{2\pi^2}\int\limits_{-\pi}^\pi
\int\limits_0^{+\infty} f\left(z,\phi\right)TL_{n}\left(z\right)
\frac{1}{z+L}\sqrt{\frac{L}{z}}\cos{\left(m\phi\right)}\;dz\;d\phi\\
\displaystyle f_{nm}^s=\frac{C}{2\pi^2}\int\limits_{-\pi}^\pi
\int\limits_0^{+\infty} f\left(z,\phi\right)TL_{n}\left(z\right)
\frac{1}{z+L}\sqrt{\frac{L}{z}}\sin{\left(m\phi\right)}\;dz\;d\phi.\label{coeffs}
\end{array}\end{equation}

  These coefficients are finite, therefore, the series in (\ref{descomp}) is $L^2$-convergent to
the function $f$ which is representing, what implies the expression
in (\ref{descomp}) is well defined.\\

\subsection{Coefficients decay}

  CHEF coefficients are uniformly bounded by
the norm of the function:
\begin{equation}
\big|f_{nm}\big|=\frac{C}{2\pi^2}\:\big|\langle
f,\phi_{nm}\rangle\big|\leq
\frac{C}{2\pi^2}\:\|f\|\cdot\|TL_{n}(r)W_{m}(\theta)\|=\|f\|
\label{acot_unif}
\end{equation}

\noindent and their sum is also bounded
\begin{equation}
\sum\limits_{m=0}^{+\infty}\sum\limits_{n=0}^{+\infty}
\big|f_{nm}\big|^2\leq\frac{C}{2\pi^2}\:\|f\|^2,\label{acot_sum}
\end{equation}

so the coefficients must tend to vanish as $n$ and
$m$ increase. It can be proved (see appendix \ref{apendice}) that the decay rate of these
CHEF coefficients is

\begin{equation}
\big|f_{nm}\big|\leq\frac{A}{|n|\;|m|^{\frac{p+1}{2}}},\hspace{0.5cm}
\forall\,n,m\in\mathbb{N},\label{acot_F-C}
\end{equation}

\noindent whenever the function $f$ is smooth enough, and $p$ is related to this smoothness; 
this ensures the quick convergence of the fits and therefore its compacteness.

\subsection{Shape parameters and fluxes}

The fact that CHEFs are radially symmetric simplifies the calculation of certain shape parameters as the 
total flux, the centroid, and the rms radius. As we will see, only the $m\in\{-2,-1,0,1,2\}$ coefficients of this
decomposition are included in the computation of these parameters.\\

  First of all, we define the integral $I_p^{n}=\int_0^R TL_{n}(r)\; r^p\;\;dr,$ with $p\geq 1$ and $R\in\mathbb{R}$. It can be proved that
$$I_p^{n}=\displaystyle 2\sum_{j=0}^{n} \left(\begin{array}{c} n \\ j \end{array}\right) (-1)^j L^{-j/2}\frac{R^{p+j/2+1}}{2p+j+2}\;Re\left(e^{in\pi/2}i^{n+j} \,_2F_1\left(n,2p+j+2,2p+j+3;\frac{-i\sqrt{R}}{\sqrt{L}}\right)\right),$$

\noindent in the case $n>0$, being $\,_2F_1$ the hypergeometric function. If $n=0$, then simply
$$I_p^{0}=\frac{R^{p+1}}{p+1}.$$

  Using this result, the flux inside a circular aperture of radius $R$ can be easily calculated from the CHEF cosine coefficients with $m=0$:
$$F(R)=\int\limits_0^{R}\int\limits_{-\pi}^\pi f(r,\theta)r\; d\theta dr=2\pi \sum_{n=0}^{+\infty} f_{n,0}^c\,I_1^{n}.$$

  In a similar way, the unweighted centroid $(x_c,y_c)$ is
$$x_c+i\,y_c=\displaystyle\frac{1}{F}\int\limits_0^{R}\int\limits_{-\pi}^\pi f(r,\theta)r^2(\cos{\theta}+i\sin{\theta})\; d\theta dr=\frac{2\pi}{F} \sum_{n=0}^{+\infty}\left[\left( f_{n,1}^c+f_{n,-1}^c\right)+i\left(f_{n,1}^s-f_{n,-1}^s\right)\right]\,I_2^{n}.$$

  Finally, the expression for the quadrupole moments $\displaystyle F_{ij}=\iint\limits_\mathbb{R} f(x,y)x_i x_j\,dx\,dy$, yields this formula for the rms radius
$$r_{rms}=\frac{F_{11}+F_{22}}{F}=\frac{2\pi}{F}\sum_{n=0}^{+\infty} f_{n,0}^c\,I_3^{n},$$

\noindent and the ellipticity
$$e=\frac{F_{11}-F_{22}+2iF_{12}}{F_{11}+F_{22}}=\frac{\sum\limits_{n=0}^{+\infty}\left( f_{n,2}^c+f_{n,2}^s\right)\,I_3^{n}}{\sum\limits_{n=0}^{+\infty} f_{n,0}^c\,I_3^{n}}.$$

  Another definition of the ellipticity, more useful as a shear estimator can be also obtained from the CHEF coefficients using the quadrupole moments \citep{melchior}:
\begin{equation}\begin{array}{@{}r@{}c@{}l}\epsilon & = & \displaystyle\frac{F_{11}-F_{22}+2iF_{12}}{F_{11}+F_{22}+2(F_{11}F_{22}-F_{12}^2)^{1/2}}=\\
& = & \displaystyle\frac{\sum\limits_{n=0}^{+\infty}\left( f_{n,2}^c+f_{n,2}^s\right)\,I_3^{n}}{2\sum\limits_{n=0}^{+\infty} f_{n,0}^c I_3^n+\sqrt{\sum\limits_{l_1,l_2=0}^{+\infty}\left[\left(f_{l_1,0}^c+\frac{1}{2}(f_{l_1,-2}^c+f_{l_1,2}^c)\right)\left(f_{l_2,0}^c-\frac{1}{2}(f_{l_2,-2}^c+f_{l_2,2}^c)\right)-\frac{1}{4}(f_{l_1,2}^s-f_{l_1,-2}^s)(f_{l_2,2}^s-f_{l_2,-2}^s)\right]I_3^{l_1}I_3^{l_2}}}.\end{array}\label{elip_melchior}\end{equation}

Using a different approach, the behaviour of the CHEF coefficients under linear transformations described in appendix \ref{apendice3}, the asymmetry parameter described by \citet{CAS} can be calculated using the CHEF coefficients:
$$A=\frac{1}{F}\int\limits_0^R \int_{-\pi}^{\pi} |f(x,y)-f^{rot}(x,y)|\;dx dy=\frac{2}{F}\iint\limits_{\mathbb{R}^2}\Bigg|\sum\limits_{m=0,\mbox{\tiny{odd}}}^{+\infty}\sum\limits_{n=0}^{+\infty}\left(f_{nm}^c \cos(m\theta)+f_{nm}^s\sin(m\theta)\right)\;TL_n(r)\Bigg| r\;d\theta dr
$$

And, finally, the S\'ersic index can be also easily expressed by means of the CHEF coefficients:
$$n=\frac{-k \ln\left(r/r_e\right)}{\ln\left(\frac{1}{C\pi\Sigma_e\e}\sum\limits_{n=0}^{+\infty}\sum\limits_{m=0}^{+\infty} \left[f_{nm}^c\cos(m\theta)+f_{nm}^s\sin(m\theta)\right]\;TL_n(r)\right)},$$

\noindent where $r_e$ is the effective radius of the galaxy, $\Sigma_e$ the surface brightness at $r_e$, and $k$ is a constant coupled to $n$ such that half of the total flux is within $r_e$.

\section{Practical implementation}

In this section we also show how to deal 
with the free parameters of the decomposition, namely the scale parameter $L$, the center $(x_c,y_c)$ 
and the total number of coefficients $N$ and $M$, something necessary for any practical application. 

\subsection{Discrete basis functions}
\label{sec_discretizacion}

  Astronomical images are formed by pixels, and therefore, any analytical basis must be 
discretized to model real data, going from the space of continuous functions defined on $\mathbb{R}^2$ to the space of matrices with size $p_1\times p_2$, that is, the size of the image being fit. 
This is achieved by integrating the CHEF basis within each pixel:

\begin{equation}
\left\{\phi_{nm}(r_j,\theta_k,L)\right\}_{nm}=\left\{\:\iint\limits_{(r_j,\theta_k) \mbox{\tiny{ pixel}}}
\frac{C}{\pi}
TL_{n}(r_j)W_{m}(\theta_k)\right\}_{nm},
\end{equation}

\noindent where $(r_j,\theta_k)$ are the coordinates of the nodes of a polar grid of size $p_1\times p_2$. 

  It is obvious that this step will not produce a complete, let alone orthonormal, basis on the pixel space. In principle, one could sample the basis at the pixel centers and fit the galaxies using an optimization scheme. The basis is so well suited to fit galaxy profiles that it can be used in this way with reasonably good results. However, completeness and orthonormality are essential for a robust and automated approach. The first property ensures the flexibility of the basis, the capability of fitting all possible galaxy profiles, including irregulars. The second is crucial for a fast implementation, since the coefficients of the decomposition can be calculated in a straightforward way, using the inner product defined in the space of real vectors. 

  Therefore, to orthonormalize our discrete basis, we chose the Modified Gram-Schmidt method (MGS), which consists in a smart rearrangement of the calculations of the Classical Gram-Schmidt process \citep{Gram}, yielding the same results as CGS in exact arithmetic, but displaying a much more stable behavior in finite-precision arithmetic. In fact, due to rounding errors, the application of the classic method produces final vectors which are often not orthogonal.  
A description of the MGS computation algorithm can be looked up in \citep{Gram}. When applied to our pixelized bases, it produces nicely orthonormal vectors, up to the larger orders tested. Using the new basis $\phi_{nm}$ we have that the image light distribution $f$ can be decomposed as 

\begin{equation}
f(r_j,\theta_k)=\sum\limits_{nm}^{NM}f_{nm}\phi_{nm}(r_j,\theta_k).
\label{discrete1}
\end{equation}

  The values of the decomposition coefficients are fast and accurately calculated by means of the usual inner product in the matrix space, that is:
\begin{equation}
f_{nm}=\sum_{j,k}f(r_j,\theta_k)\phi_{nm}(r_j,\theta_k),
\label{discrete2}
\end{equation}
where the sum $\sum_{j,k}$ goes over all the pixels in the image. 

 The orthonormalization method needs to be applied only once, for a given scale and image size. Given that the CHEF reconstruction is relatively robust to the scale $L$ and image size, it is possible to produce a grid of basis functions, with different scales and sizes and precalculate the orthonormalization in advance. This significantly improves the efficiency of the method when applied to large data volumes. 

\subsection{Number of coefficients, center and scale size}\label{centro_escala}

As in the case of any similar method, the decomposition has several free parameters: the scale $L$, the total number of coefficients $(N+1)\cdot(2M+1)$ and the origin $(x_c,y_c)$. It is obvious that if we grossly err in the choice of one of them, the algorithm will not work properly. Conveniently, quasi-optimal choices for all these parameters, except $N$ and $M$, can be obtained directly from the output of SExtractor.  

  The center of the reconstruction is a sensitive parameter, since the basis functions have an extreme at its position: it should match the maximum of the light distribution of the object, since a wrongly determined center would require an innecessarily high number of coefficients for an accurate fit. Luckly the centroid calculated by SExtractor provides a suitable center in practically all cases. 

  The scale parameter $L$ is not related to the shape of the CHEF basis, but to their width (the scale has an ``accordion'' effect over the basis functions). The larger the scale is, the slower the basis functions reach their last extreme, as can be seen in figure \ref{fig1}. A scale size which is too small would overfit the high frequency noise in the central part of the image, whereas a scale size which is too large would smooth over the galaxies, incapable of fitting the central peak and the tiny features. In principle one could perform the fits iteratively, looking for the best scale $L$. However, we have verified that the quality of the fit is robust with respect to the value of $L$ and using the SExtractor half-light radius estimate as a proxy for its value yields extremely accurate results. It  is therefore not necessary to optimize this parameter.   

  The remaining parameter, the orders of the decomposition $N$ and $M$ need to be carefully selected. We must choose a number of coefficients which is large enough to accurately model the image, but without overfitting the noise. The number of coefficients is thus optimized until the reconstruction residual approaches 1:

\begin{equation}
\chi^2=\frac{1}{n_{pixel}-n_{coeff}}\displaystyle\sum_{j,k}
\frac{
\left(f(r_j,\theta_k)-\sum\limits_{nm}f_{nm}\phi(r_j,\theta_k,L)\right)^2}
{\delta^2(r_j,\theta_k)}
\end{equation}

where $f$ is the observed image, $f_{nm}$ are its CHEF coefficients, 
$\delta^2$ is referred to the pixel noise map, $n_{pixel}$ is the number of pixels 
in $f$ (that is, $n_{pixel}=p_1\cdot p_2$), and $n_{coeff}$ represents the number of CHEF coefficients used to 
decompose $f$ (i.e., $n_{coeff}=(N+1)(2M+1)$). At the point where $\chi^2=1$, the residual between the observed 
data and the model is consistent with the noise level. 
\\

\subsection{The CHEF pipeline}
\label{sec_practica}

\noindent The process to decompose real image data using the CHEF basis has been implemented in a Python software package, which is being applied to the ALHAMBRA survey data \citep{moles,benitez}. In the process outlined below we do not deal separately with the PSFs, fitting directly the objects without attempting any deconvolution. That will be the subject of a separate paper. 

\begin{enumerate}
\item \textit{Image analysis with SExtractor and source filtering}. We run SExtractor on the image, to select galaxies and 
estimate the parameters we need for the fit, namely the centroid and the half-light radius, which will be used to determine the origin 
and the scale size of the CHEFs, respectively. 

\item \textit{Choice of the reconstruction size and the maximum number of coefficients}. We have found that a good choice for the size of the reconstruction is to set $L_{max}=4\times r_{hl}$, where $r_{hl}$ is the half-light radius of the galaxy, as measured by SExtractor. In practically all cases this fully includes the galaxy flux. 

   As a general rule, the larger the number of pixels in the image to be fit, the larger the required number of coefficients. For instance, for two images of the same galaxy, the one with a smaller pixel scale will require more coefficients (assuming of course that the S/N is similar in both cases). We set $N^{max}, M^{max}$ and choose this value according to the image size. For instance, for a regular image of $60\times 60$ pixels the usual choice is $N^{max}=M^{max}=7$, whereas for more extended sources of about $120\times 120$ pixels we set $N^{max}=M^{max}=10$.

 Once the image size is determined, we mask out all the pixels belonging to other objects present within the frame. 

\item \textit{Evaluation of the basis}. The CHEF basis corresponding 
to $N^{max}$ and $M^{max}$ is evaluated and orthonormalized using the modified 
Gram-Schmidt algorithm (see section \ref{sec_discretizacion}). Thanks to the robustness 
of the CHEF decomposition, it is possible to precalculate this step, for a grid of different scale parameters $L$, so that the software just has to look for the closest value stored in the database, making it computationally much faster. For large data volumes this results in a speed up of more than a factor of 3.   

\item \textit{Determination of the optimal number of coefficients}. As explained in Sec. \ref{centro_escala}, we choose the optimal number of coefficients $N$ and $M$ iteratively, by choosing the values which yield  $\chi^2\approx 1$. 
The corresponding basis functions, up to the required order, are extracted from the whole orthonormalized set previously 
computed in step 3.

\item \textit{Evaluation of CHEF coefficients and calculation of the model}. The 
$n_{coeff}=(N+1)\times(2M+1)$ coefficients of the CHEF expansion and the resulting model are calculated by simple scalar products, 
as described in equations (\ref{discrete1}) and (\ref{discrete2}). 

\item \textit{Iteration of the algorithm for overlapping objects}. For those objects whose flux may be affected by nearby objects we repeat steps 2-5, but  instead of masking out the objects, we subtract their profiles already calculated in the previous steps. 

\end{enumerate}

\begin{figure*}%[h]
\begin{center}
\includegraphics[width=15cm]{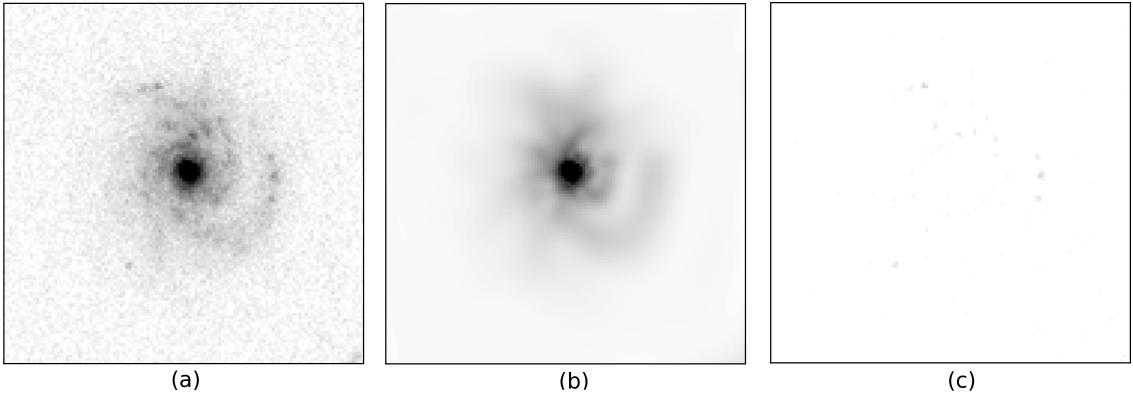}
\caption{CHEF decomposition of a spiral galaxy from the UDF:
a) original image, b) CHEF reconstruction
using $N=15$ radial and $M=14$ angular coefficients, (c) residuals}
\label{main_cheby}\end{center}
\end{figure*}

\begin{figure*}[t]
\begin{center}
\includegraphics[width=17cm]{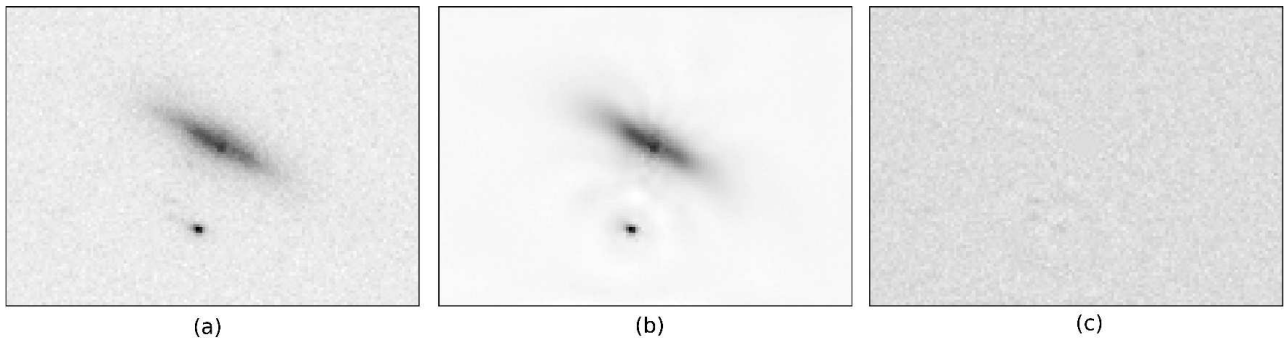}
\caption{CHEF decomposition of an edge-on galaxy from the UDF: a) original image b) CHEF reconstruction with $N=5$ and $M=7$, and c) residuals 
(note that the small object below the main galaxy is also fit separately)
}\label{gal_edgeon}\end{center}
\end{figure*}

\begin{figure*}
\begin{center}
\includegraphics[width=15cm]{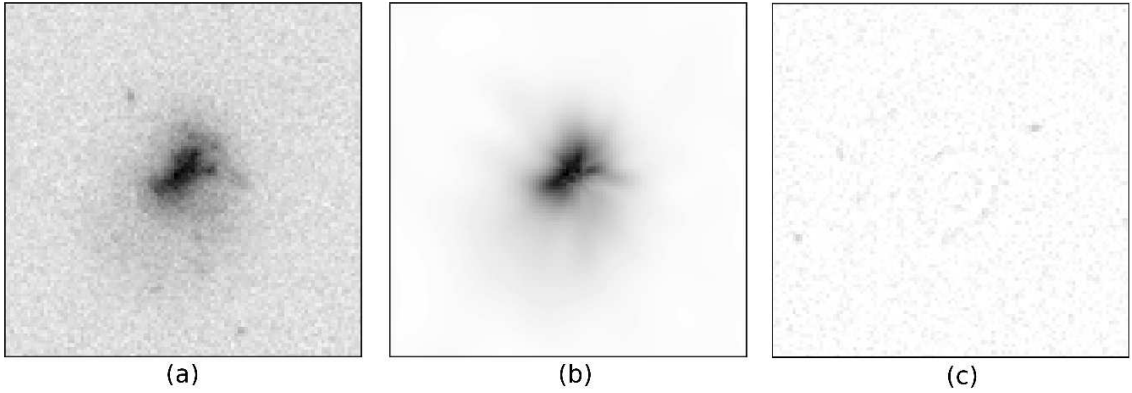}
\caption{CHEF decomposition of an irregular galaxy from the UDF: a) original image b) CHEF reconstruction with $N=5$ and $M=7$, and c) residuals}\label{gal_irregular}\end{center}
\end{figure*}

  Figures \ref{main_cheby},\ref{gal_edgeon}, and \ref{gal_irregular} show some examples of Hubble Ultra Deep Field \citep{UDF,coe} galaxies. The large amount of detail revealed in those images makes them a useful benchmark for galaxy decomposition methods. These objects, while typical, are not trivial to fit due to the complicated structure revealed by the high $S/N$ observations. The first is a spiral galaxy, displayed in figure \ref{main_cheby}. The image frame has $121\times 121$ pixels, and the algorithm establishes that the best fit is obtained with $N=15$ and $M=14$, i.e., using $n_{coeff}=(15+1)(14\times 2+1)=464$ coefficients. The very low level of the residuals shows that the CHEF basis can easily fit both the central bulge and large structures like spiral arms. We apply the same procedure to other two galaxies, an edge on galaxy with very high observed ellipticity (Fig. \ref{gal_edgeon}) and a highly irregular galaxy (Fig. \ref{gal_irregular}). The fits are also excellent, what illustrates the flexibility of the CHEF basis, able to handle most galaxy morphologies.  

\section{Comparison with other methods}
\label{comparison}

   We compare the CHEF decomposition with two of the most widely available techniques for galaxy modeling, GALFIT \citep{peng} and the IDL Shapelets software. GALFIT \citep{peng} uses several two-dimensional analytic models to fit the galaxy images, as the S\'ersic profile, the exponential disk, the Nuker law or the Gaussian and Moffat/Lorentzian functions. It is possible to use an arbitrary number of components. The shapelets software \citep{massey} is based on an orthonormal basis built from Hermite polynomials and Gaussians.   Figure \ref{muestra} shows the results of fitting 12 randomly chosen galaxies from UDF with GALFIT, IDL Shapelets and our CHEF pipeline. 

 The GALFIT results show that radially symmetrical multicomponent models fail to adequately represent the wide variety of shapes displayed by galaxies observed at high S/N, and that it is essential to include angular components, as indeed the new version of GALFIT does. However, an essential problem of GALFIT, and in general of any similar approach, is that by using a combination of several analytical profiles, GALFIT is mathematically equivalent to a decomposition into a set of functions which not only do not form an orthogonal basis, but are highly degenerate among themselves. Therefore as soon as the number of components increases, the fits must be performed interactively, in order to provide the right initial parameters, the number and type of the components, etc. This makes very difficult its use to obtain automatic, highly accurate fits of large numbers of galaxies. 

  We have run the IDL shapelets software allowing $n_{max}=20$ as the maximum order, equivalent to  $n_{coeff}=400$ coefficients. For comparison, we use $N^{max}=M^{max}=10$ , equivalent to $n_{coeff}=231$ coefficients, for the CHEF decomposition, which also runs faster, even taking into account the orthonormalization process. Despite of using more coefficients, we see the shapelet reconstruction often fails to adequately fit all galaxy shapes and tends to produce  ring-like artifacts, probably as a consequence of the inestability of the radial profile fits \citep{jim}. 

  The CHEF decomposition manages to keep the residuals to a very low level, fitting very complicated galaxy shapes, producing very few artifacts. In addition, as we will see below, it obtains highly accurate estimates of galaxy total fluxes and shapes. 

\begin{figure*}
\centerline{\includegraphics[width=10cm]{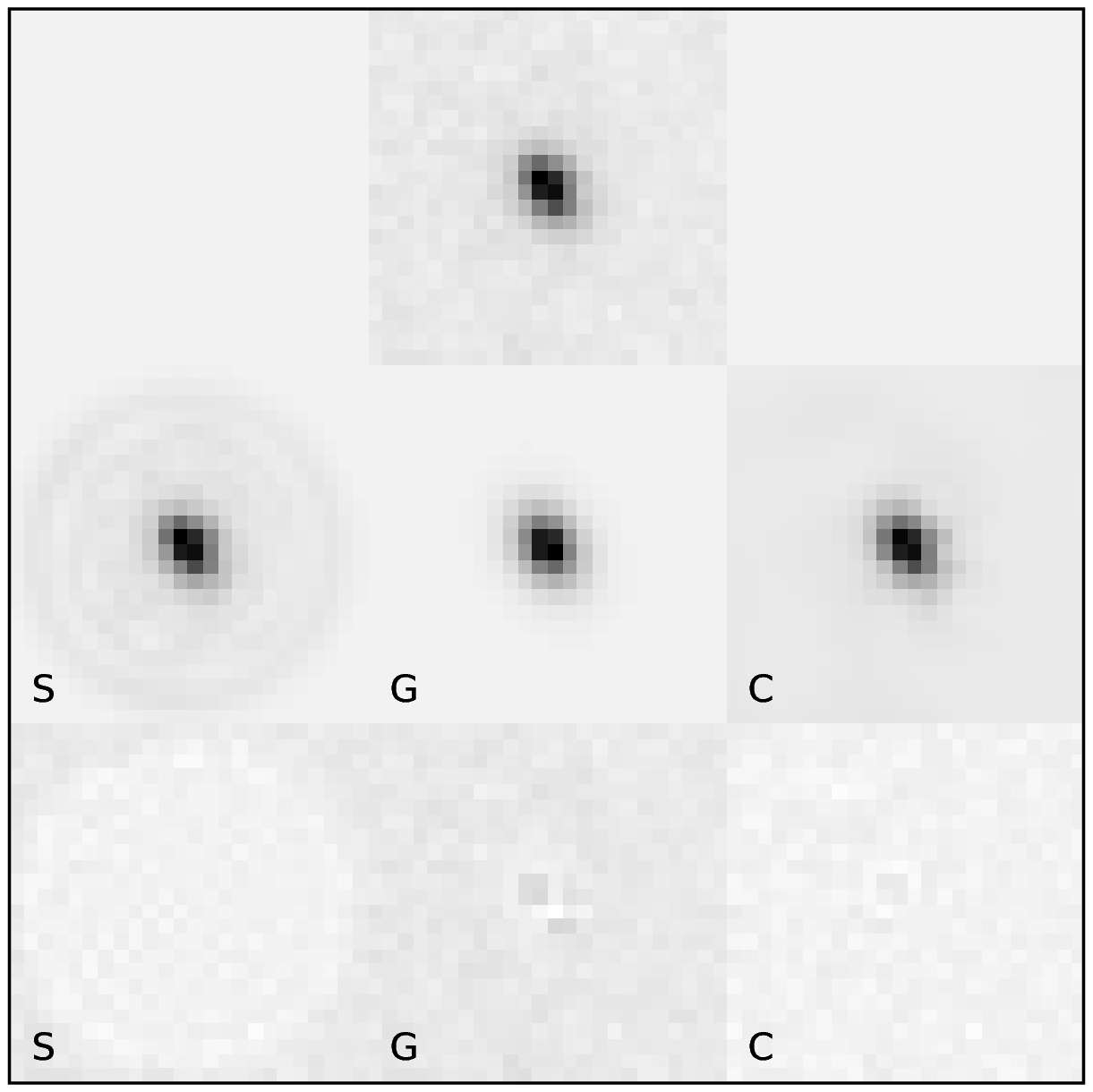}\includegraphics[width=10cm]{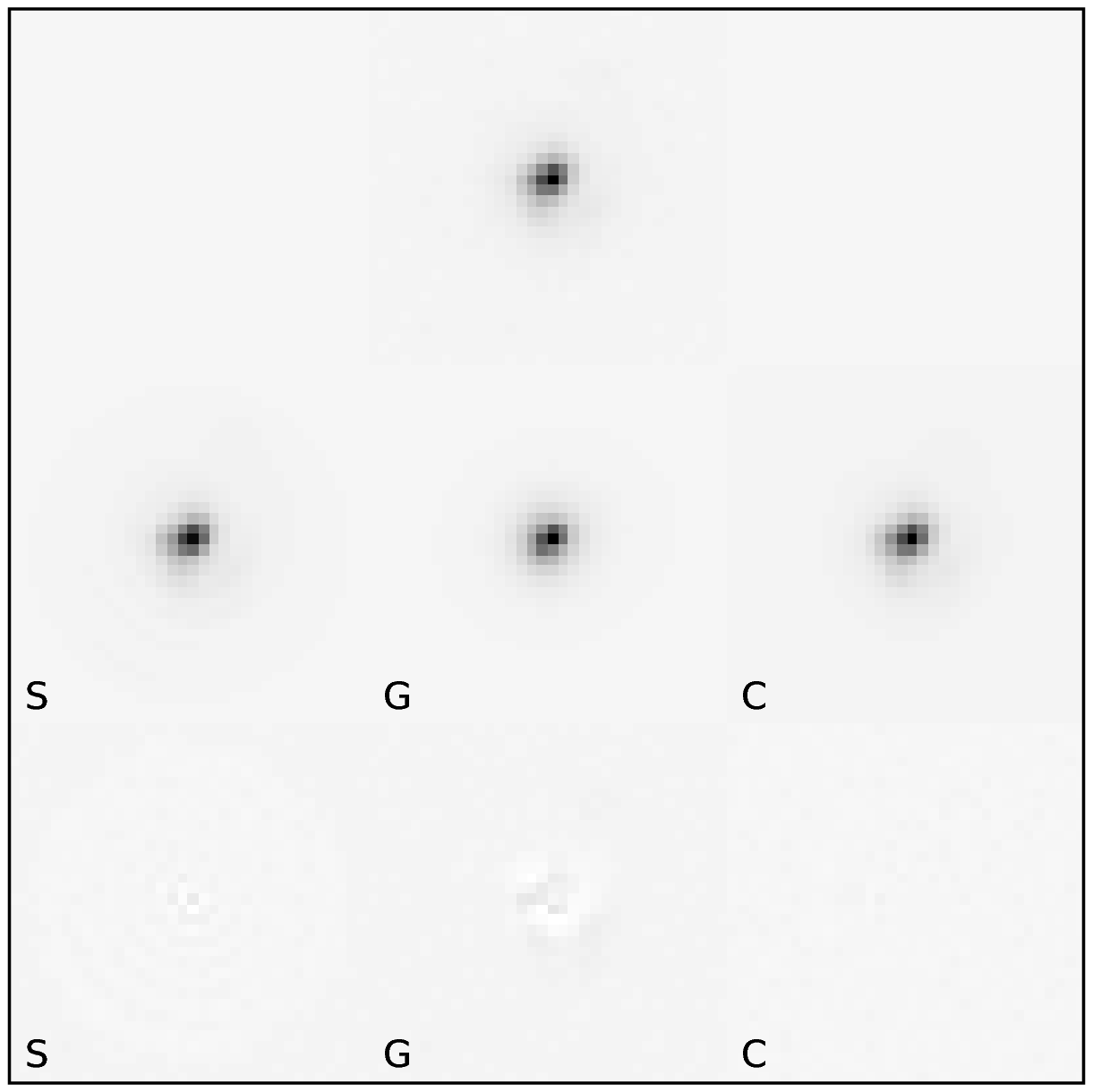}}
\centerline{\includegraphics[width=10cm]{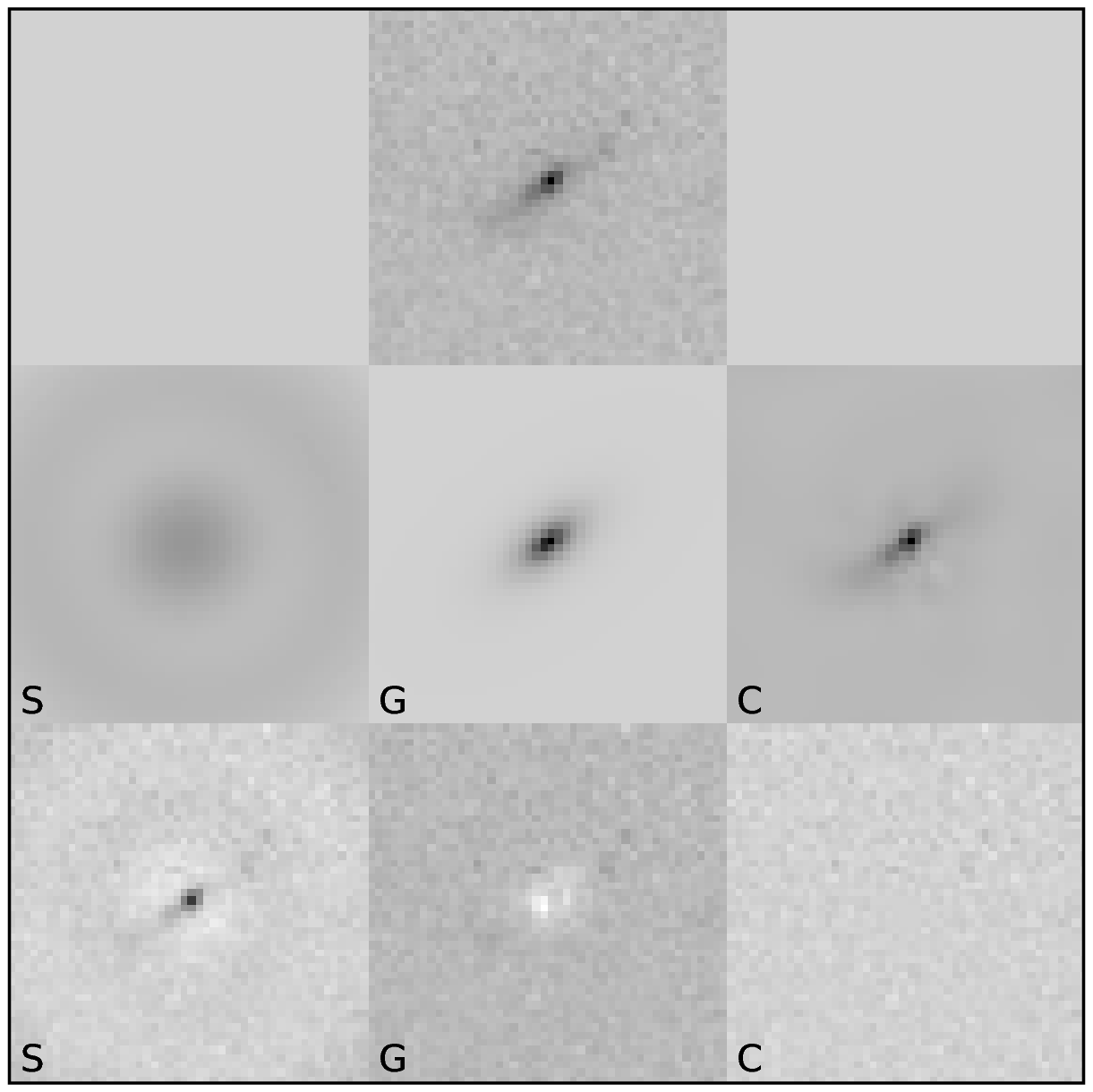}\includegraphics[width=10cm]{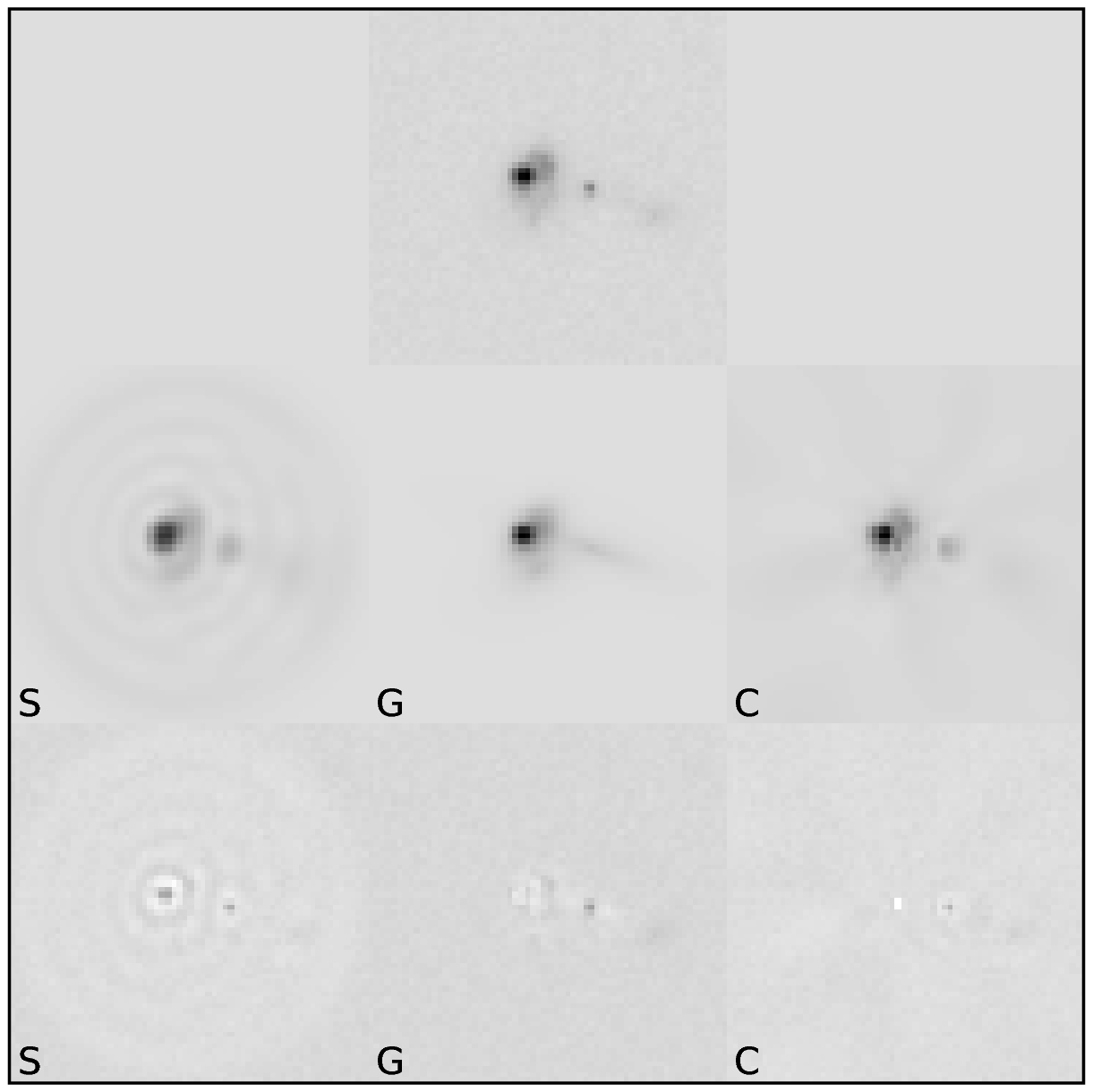}}
\centerline{\includegraphics[width=10cm]{4_total.eps}\includegraphics[width=10cm]{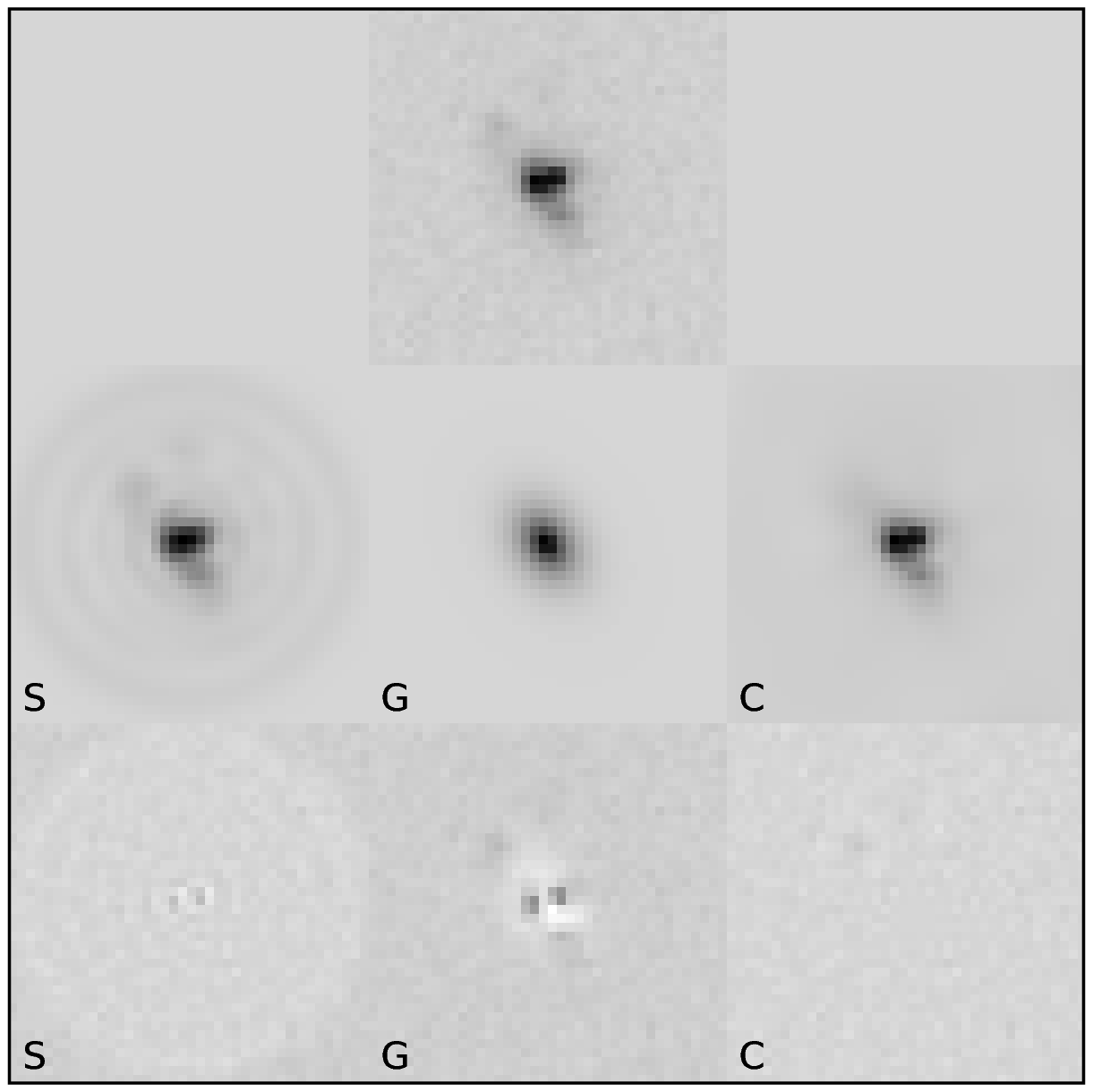}}
\end{figure*}
\clearpage

\begin{figure*}
\centerline{\includegraphics[width=10cm]{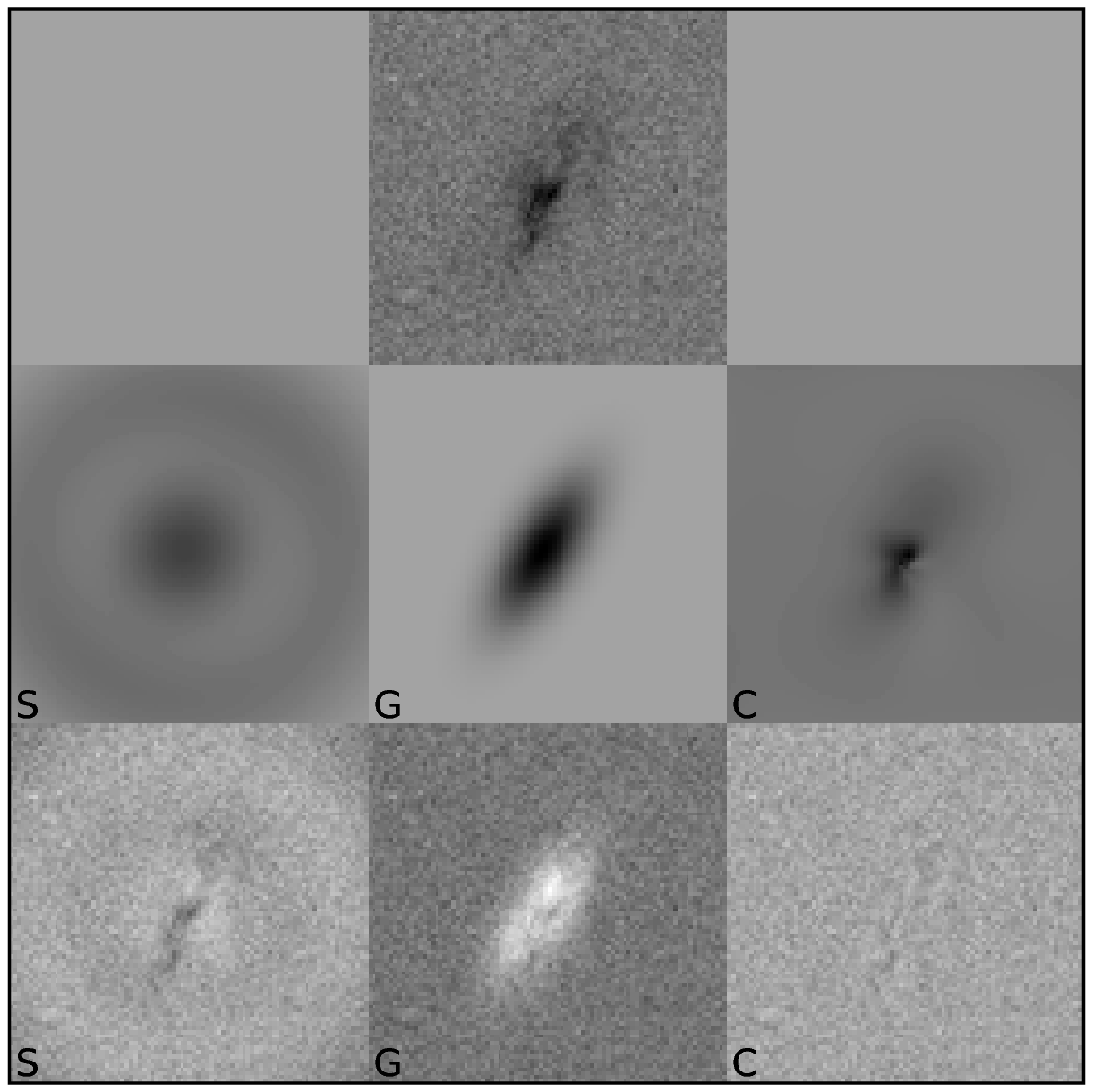}\includegraphics[width=10cm]{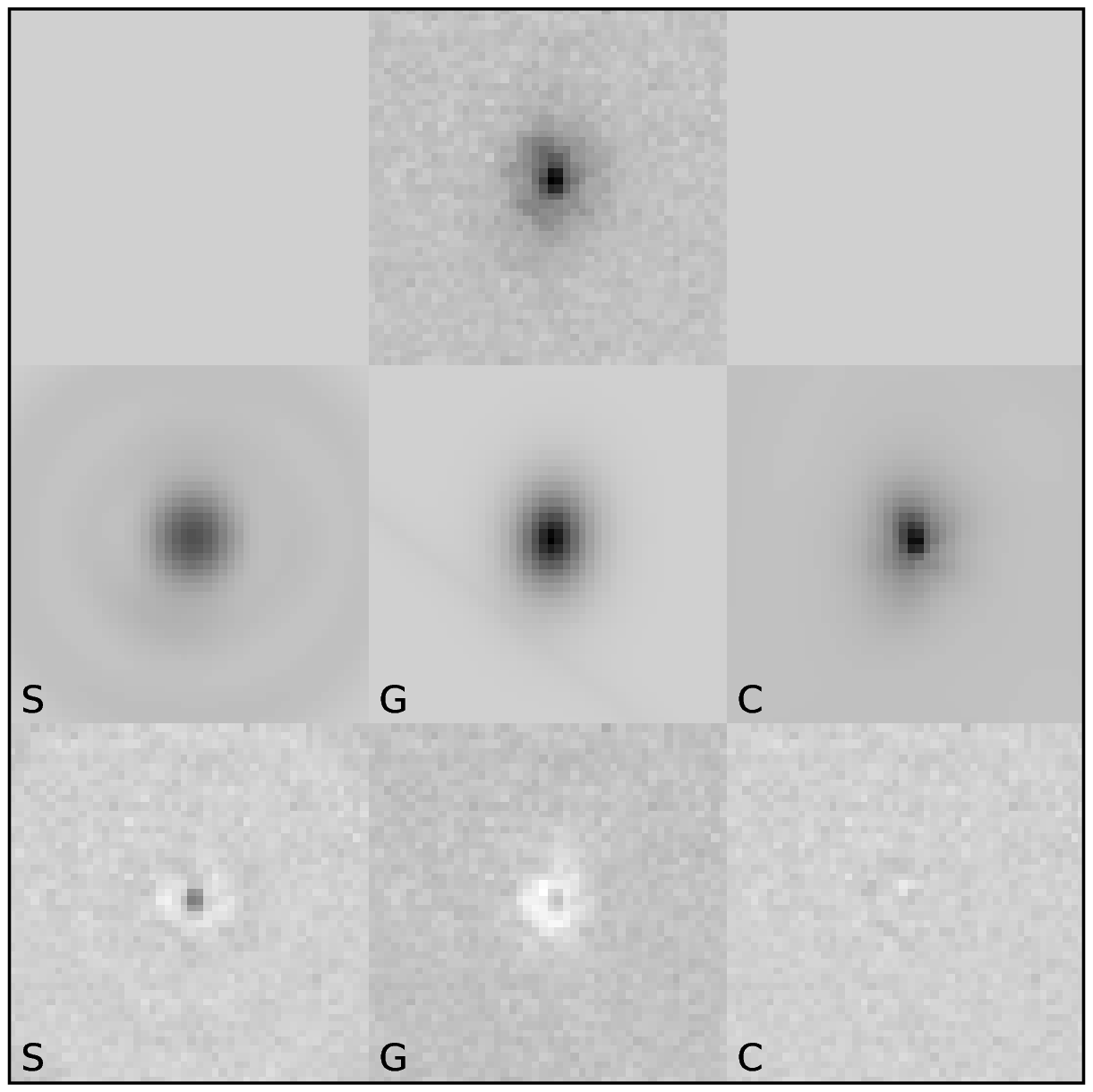}}
\centerline{\includegraphics[width=10cm]{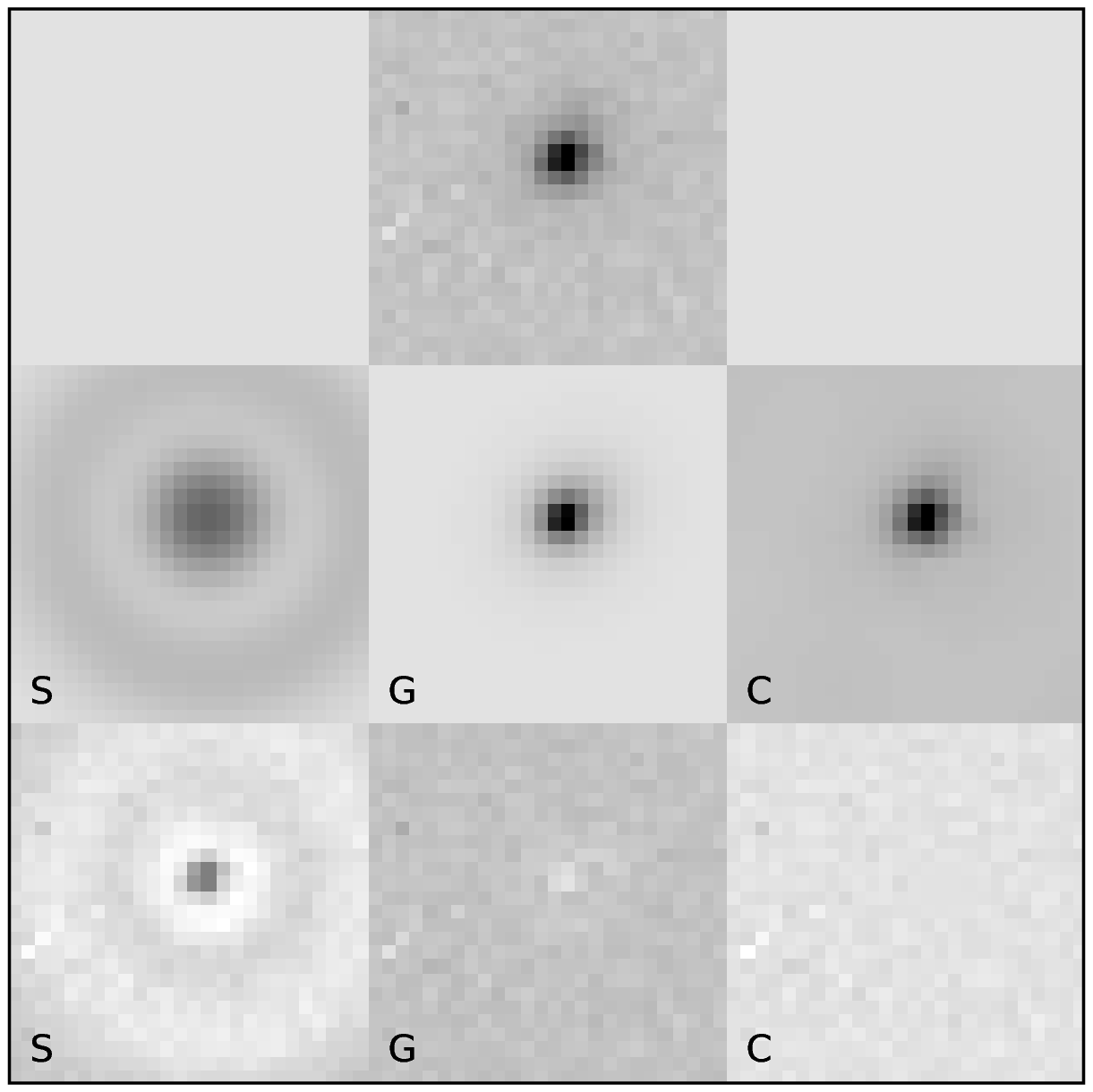}\includegraphics[width=10cm]{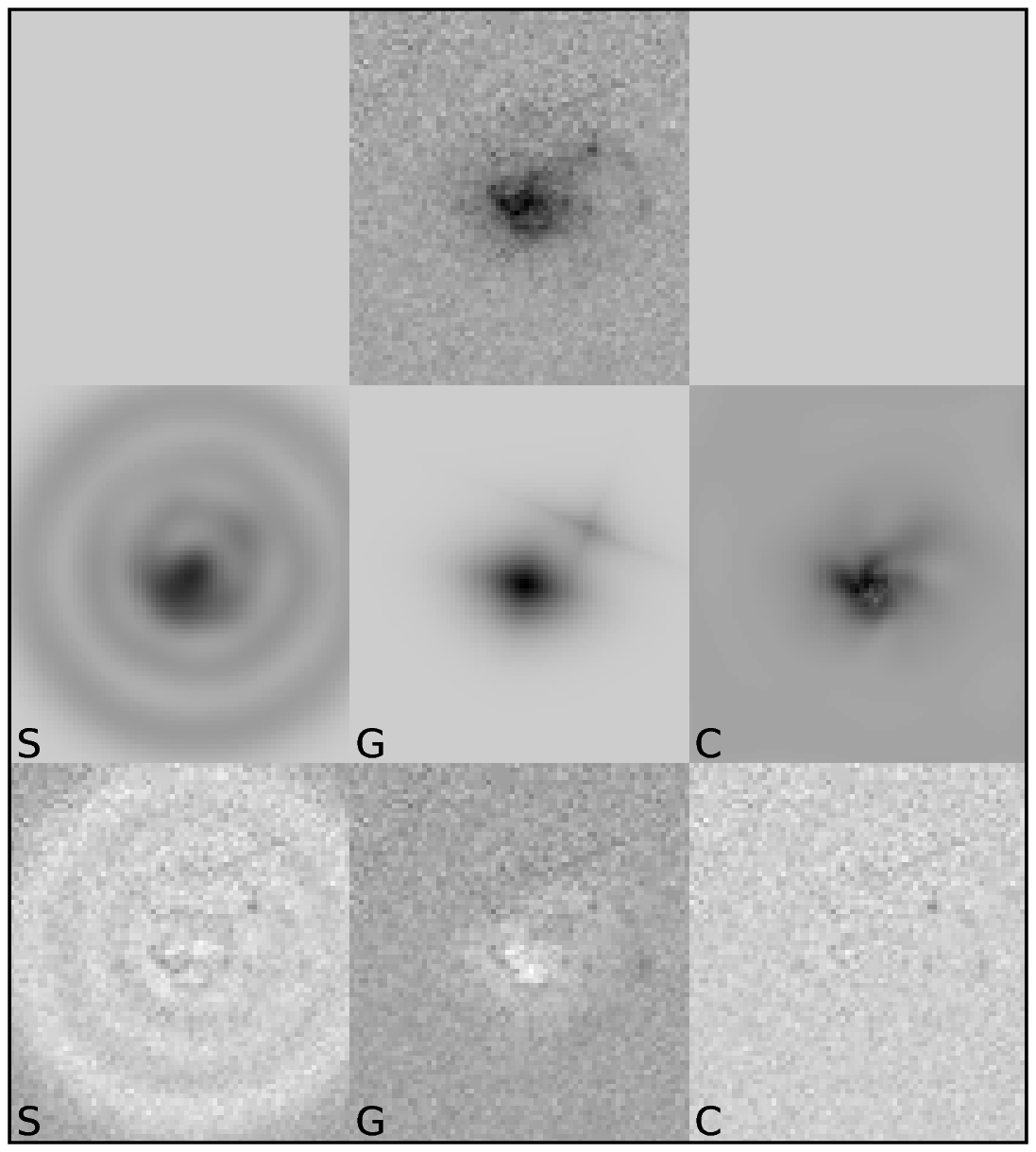}}
\centerline{\includegraphics[width=10cm]{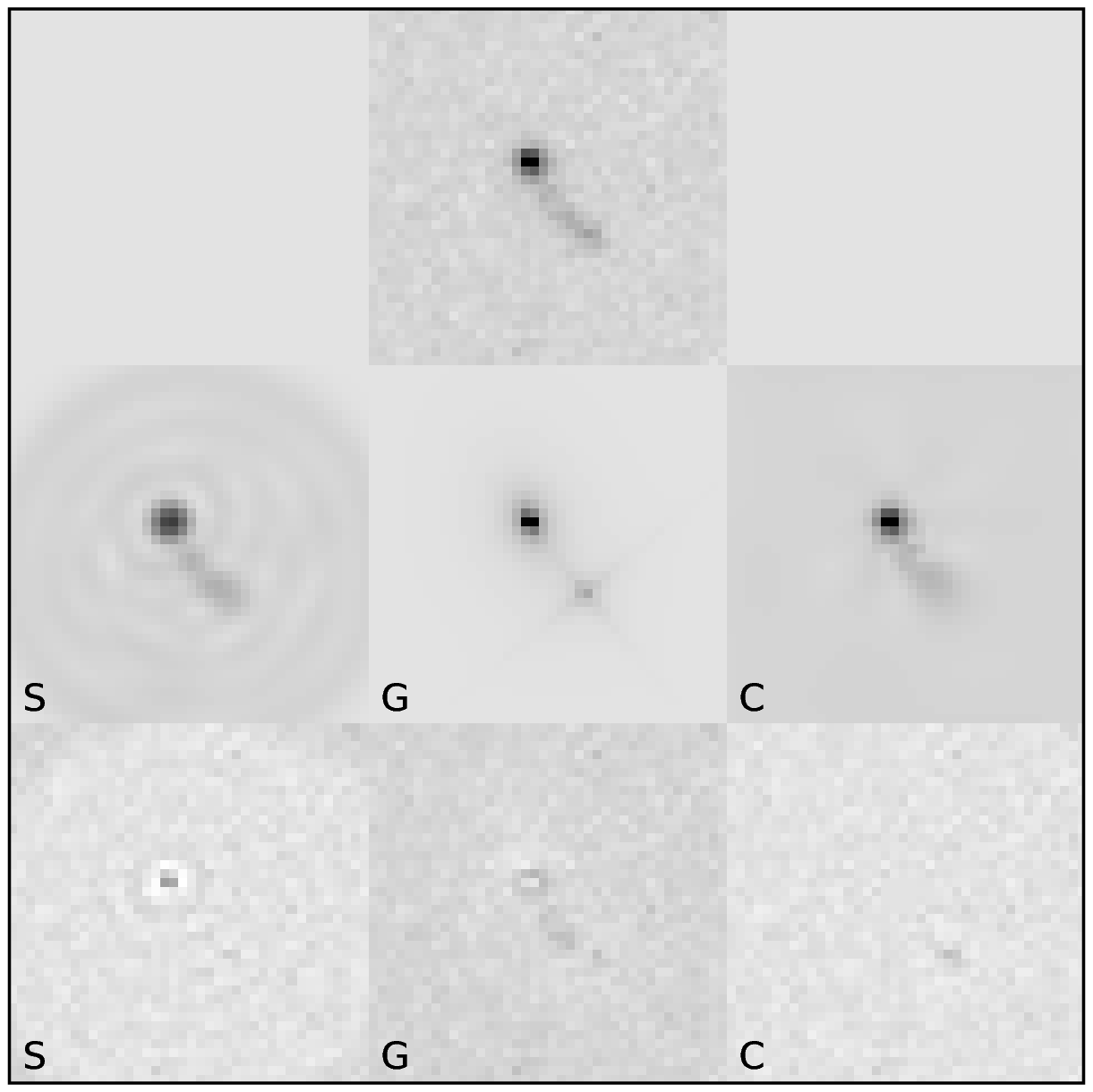}\includegraphics[width=10cm]{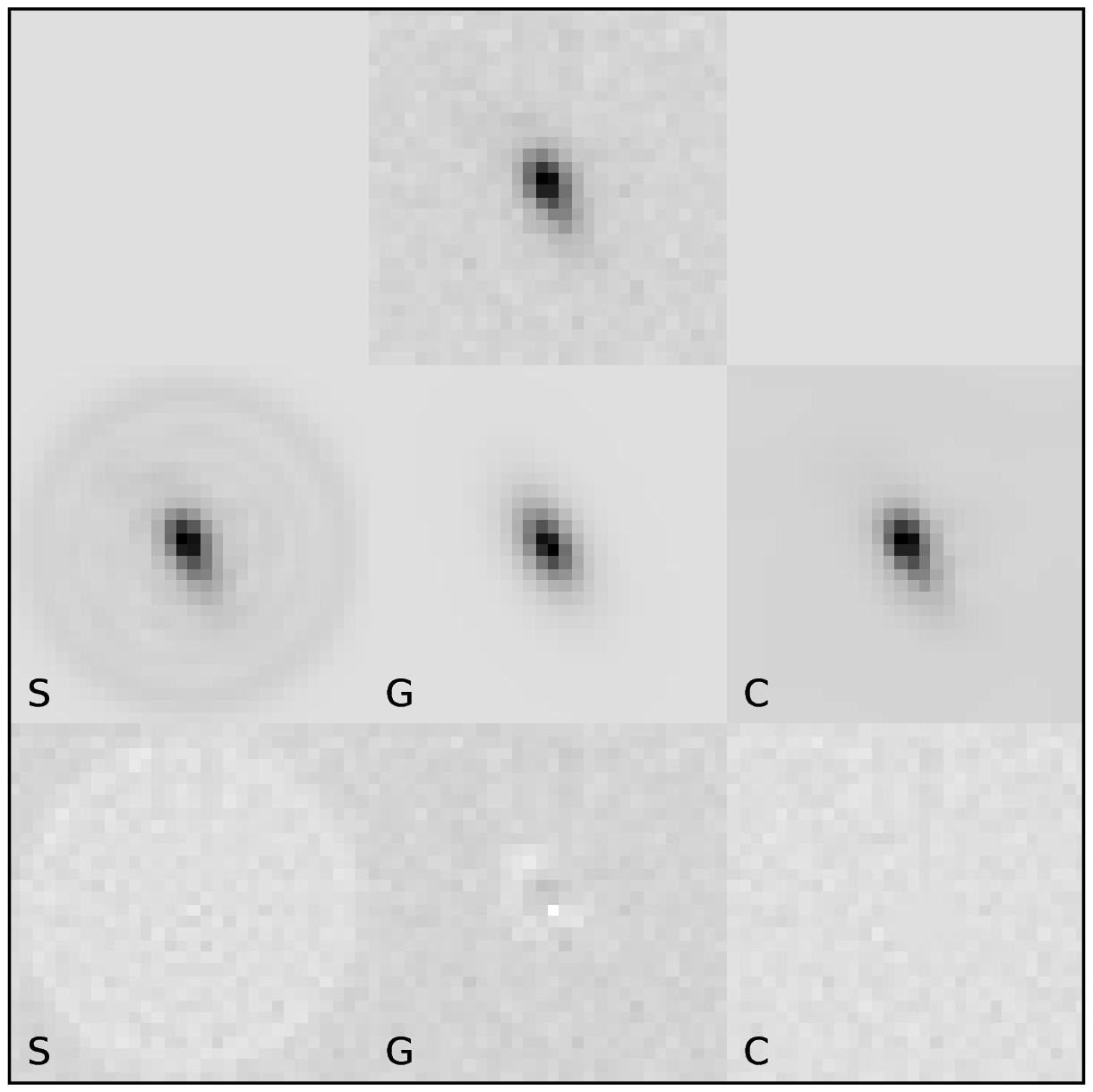}}
\caption{Reconstructions and obtained residuals obtained after applying the three methods to an ensemble of galaxies randomly extracted from UDF: images on the top correspond to the original galaxies, first row below it represents the reconstructions obtained by shapelets, GALFIT and CHEF softwares (from left to right), and second row shows the resulting residuals, in linear scale.}\label{muestra}
\end{figure*}
\clearpage

\subsection{Magnitudes and ellipticities}
 
  As stated above, the main motivation to develop the CHEF basis is having a tool to 
automatically measure accurate fluxes and shapes in the context of large galaxy surveys. 

  Here we test the performance of CHEFs using an ensemble of 350 mock galaxy profiles, 
simulated using S\'ersic profiles with indices ranging from 0.5 to 4, and sheared by 
different amounts, ranging from ellipticity 0 to 0.5, by applying to the cartesian coordinates the following transformation:

$$\left(\begin{array}{c}x^s\\y^s\end{array}\right)=\left(\begin{array}{cc} 1-\gamma_1 & -\gamma_2\\-\gamma_2 & 1+\gamma_1\end{array}\right)\left(\begin{array}{c}x\\y\end{array}\right).$$

\begin{figure*}
\begin{center}\includegraphics[width=17cm]{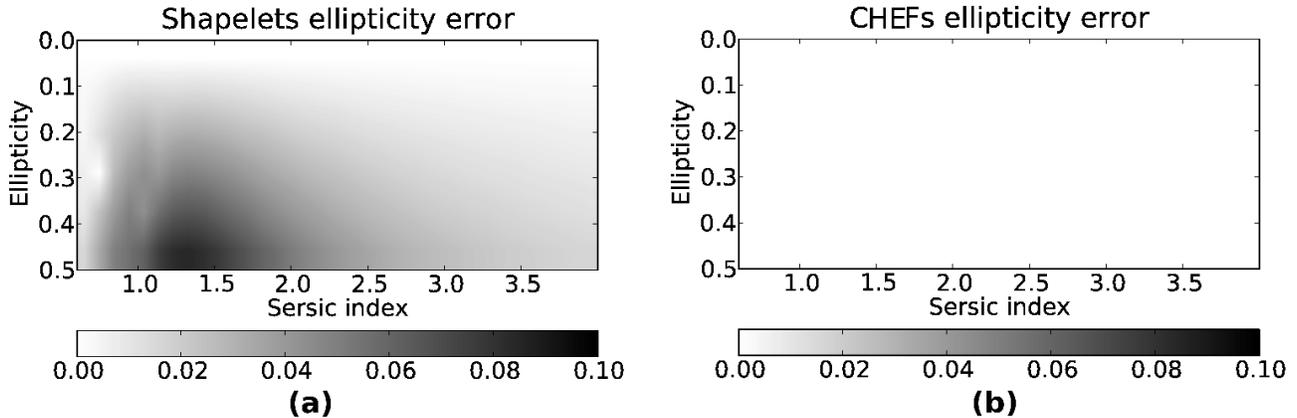}
\end{center}\caption{Ellipticity absolute error obtained by applying shapelet and CHEF algorithms to a sample of simulated galaxies with different S\'ersic profiles and ellipticity parameters.}\label{ellipt_comparison}
\end{figure*}

 We then fit the galaxy shapes using the IDL shapelet software and our CHEF pipeline. We can see in Fig \ref{ellipt_comparison} that, as expected, shapelets fail to adequately objects with high ellipticities, specially for S\'ersic indexes $n\approx 1.2$, where the relative error in the measurement of the ellipticity reaches $10\%$. Although we are using circular shapelets here, it seems that elliptical shapelets are also affected by this problem \citep{jim}. 

 Using the formula (\ref{elip_melchior}), we estimate the ellipticities of our CHEF fits. Our error is $\lesssim 0.5\%$ for all the explored combinations of profiles and shears. This shows that CHEFs are very promising as a tool for weak lensing measurements, and issue which will be explored elsewhere. 

\begin{figure*}
\begin{center}\includegraphics[width=17cm]{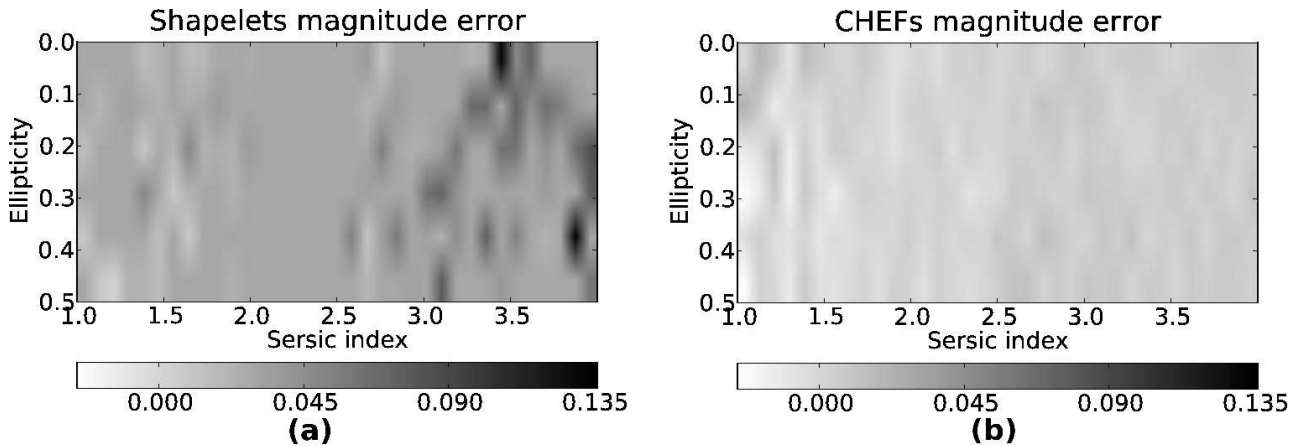}
\end{center}\caption{Magnitude relative error obtained by applying shapelet and CHEF algorithms to a sample of simulated galaxies with different S\'ersic profiles and ellipticity parameters.}\label{magnitude_comparison}
\end{figure*}

  To verify the performance of the CHEFs for photometric purposes, we added Gaussian noise to the galaxies above, and applied both the shapelets and CHEF algorithms to the data. The results are presented in  Fig \ref{magnitude_comparison}. Again we see that CHEFs provide highly accurate measurements of the total flux for almost all combinations of profile steepness and ellipticity, with a typical error of only $\approx 0.4\%$. Shapelets, on the other hand, introduce up to $4\%$ errors in this very high S/N simulated images. 

\section{Conclusions}

  Motivated by the need to perform accurate galaxy photometry and shape measurements in large surveys like the upcoming JPAS \citep{jpas}, we have developed a new orthonormal basis formed by a combination of rational Chebyshev and trigonometric functions for, respectively, the radial and angular components, and that we called CHEF functions, or CHEFs. 

  The new basis displays remarkable flexibility, being able to accurately fit all kinds of galaxy shapes, including irregulars, spirals, ellipticals, highly compact and highly sheared galaxies. It does this while using fewer components that similar methods, as shapelets, and without producing artifacts, due to the efficiency of the rational 
Chebyshev polynomials to fit quickly decaying functions like galaxy profiles. The method is lineal as well as very stable and robust, therefore, able to work very fast and to automatically process large numbers of galaxies. 

 Due to the high quality of the fits in the central parts of the galaxies, and the ability of the CHEF basis to compactly model galaxy profiles up to very large distances, the method provides highly accurate estimates of total galaxy fluxes and ellipticities. 

Future work will explore in more detail the application of the method to problems like 
multiband photometry, weak shear measurements, deconvolution, morphological classification, etc. 

\acknowledgments
The authors thank Renato \'Alvarez for reading an early draft and making useful comments. 
We also had interesting and useful conversations with Sarah Bridle, Peter Melchior and 
Joel Berg\`e. We acknowledge the support and help received during the visits to Jet Propulsion Laboratory, University College of London and Space Telescope Science Institute. This work was made possible by the financial support of the Consejo Superior de Investigaciones Cient\`ificas and its I3P grants program, the ALHAMBRA grant AYA2006-14056 and the government project AYA2010-22111-C03-00.

\appendix

\section{Coefficients decay} \label{apendice}

Let us calculate the decay rate of the CHEF
coefficients using the fact that a Chebyshev series is a Fourier
series with a change of variable, in such a way that the Chebyshev
coefficients of a univariate function $g$ are proportional to the
Fourier coefficients of the composition $g\circ\cos$:
\begin{equation}
g_n=2\,\widehat{G}[n],\hspace{0.5cm}\forall\,n\in\mathbb{N},\hspace{0.5cm}
\mbox{with}\;\;G(t)=g(\cos(t)).\label{cosine_series}
\end{equation}

\prop Let $f$ be a function of the space ${L}^1([-1,1])$.
Let us define the function $F$ in the following way:
\begin{equation}\begin{array}{rccc}F: & [-\pi,\pi] & \rightarrow & \mathbb{R} \\
& t & \mapsto & f(\cos t)\end{array}\end{equation} \noindent Then,
the Fourier coefficients of $F$ are proportional to the Chebyshev
coefficients of $f$, that is,
\begin{equation}\widehat{F}[n]=\frac{1}{k}\,f_n, \hspace{0.5cm}\forall\,
n\in\mathbb{N}\label{eqprop1}\end{equation} \noindent where $f_n$ is the n-th
coefficient in the Chebyshev series of $f$ and $\displaystyle
k=\left\{\begin{array}{ll}
k=1, & \mbox{if }\;n=0\\ k=2, & \mbox{if }\;n>0\end{array}\right.$.\label{prop1}\\

\begin{demo} \rm
The function $F$ is a $2\pi$-periodic function, so its Fourier
series can be calculated:
$$\begin{array}{r@{}c@{}l} \widehat{F}[n] & = &
\displaystyle\frac{1}{2\pi}\int\limits_{-\pi}^\pi
F(t)\e^{-int}\,dt=\frac{1}{2\pi}\int\limits_{-\pi}^0
F(t)\e^{-int}\,dt+ \frac{1}{2\pi}\int\limits_0^\pi
F(t)\e^{-int}\,dt=\\
& = & \displaystyle \frac{1}{2\pi}\int\limits_0^\pi
F(t)\left[\e^{int}+\e^{-int}\right]\,dt=
\frac{1}{\pi}\int\limits_0^\pi
F(t)\cos(nt)\,dt=\\
& = & \displaystyle\frac{1}{\pi}\int\limits_{-1}^1
f(x)\cos(n\cdot\arccos
x)(1-x^2)^{-1/2}\,dx=\frac{1}{\pi}\int\limits_{-1}^1
f(x)T_n(x)(1-x^2)^{-1/2}\,dx=\frac{1}{k}\,f_n.
\end{array}$$
\end{demo}
\rm
 This result is extremely useful since it allows us to
transfer the Fourier series properties to the CHEF
one. In addition to this, it is well-known the Fourier coefficients of a smooth 
enough function $f$ are bounded, as the following theorem states.

\thm \citep{cizek} Let $f$ be a periodic function with an interval
of periodicity $[0,P)$, continuous with its derivatives up to
order $p-1$, $p\geq1$, that is, $f\in\mathcal{C}^{p-1}$. Let its
derivative of order $p$ be piecewise continuous in the interval.
Then, there exists a constant $A>0$ such that, for Fourier
coefficients of the function $f$, we have
\begin{equation}
\big|\hat{f}[k]\big|\leq\frac{A}{|k|^{p+1}},
\hspace{0.5cm}\forall\,k\in\mathbb{N}.\label{eqth1}
\end{equation}\label{th1}
\rm
\noindent Therefore, using (\ref{eqprop1}) and
(\ref{eqth1}) it is easy to reach the decay rate of the
CHEF coefficients, as it is shown in the following
proposition.

\prop Let $f$ be a function of the space
$\mathcal{L}^2([0,+\infty]\times[-\pi,\pi])$ defined in polar
coordinates. Let us assume that $f$ is continuous in the angular
coordinate with its partial derivatives up to order $p-2$, and
that its partial derivative of order $p-1$ is piecewise continuous
in the interval $[-\pi,\pi]$. Let $f$ be also piecewise continuous
in the azimuthal coordinate (the interval $[0,+\infty))$. Then,
there exists a constant $C>0$ such that for the CHEF
coefficients of $f$ it is had the following boundary:
\begin{equation}\displaystyle|f_{nm}|\leq\frac{C}{|n|\,|m|^{\frac{p+1}{2}}},
\hspace{0.5cm}\forall\,n,m\in\mathbb{N}.\end{equation}

\begin{demo}\rm
Let us define the function $g$ as
$$\begin{array}{rccc}g: & [-\pi,\pi] &
\rightarrow & \mathbb{R} \\ & \theta & \mapsto & \displaystyle
\int\limits_0^\infty f(r,\theta)\,TL_n(r)\,\frac{1}{r+L}\sqrt{
\frac{L}{r}}\,dr\end{array}.$$

This function can be extended in a periodic way it
fulfils that it is continuous with its derivatives up to order
$p-1$, as well as the p-th derivative is piecewise continuous in
the periodicity interval $[-\pi,\pi]$. Then, using the theorem
\ref{th1}, there exists a non-negative constant $A$ such that
$$|\hat{g}[l]|\leq\frac{A}{|l|^{p+1}},\hspace{0.5cm}\forall\,l\in\mathbb{N}.$$

Moreover, there exists a relationship between the
CHEF coefficients of $f$ and these Fourier
coefficients of $g$:
$$f_{nm}=\int\limits_0^\infty\int\limits_{-\pi}^\pi
f(r,\theta)\,TL_{n}(r)\e^{-im\theta}\frac{1}{r+L}\sqrt{
\frac{L}{r}}\,dr=\int\limits_{-\pi}^\pi g(\theta) \e^{-im\theta}
d\theta=2\pi\hat{g}[m].$$

Therefore,
\begin{equation}\displaystyle
|f_{nm}|\leq\frac{A}{|m|^{p+1}}.\label{acot1}\end{equation}

On another front, let us define a function $h$ as
$$\begin{array}{rccc}h: & [-\pi,\pi] &
\rightarrow & \mathbb{R} \\ & t & \mapsto & \displaystyle
\int\limits_{-\pi}^\pi f\left(\frac{1+t}{1-t}\,L,\theta\right)
\e^{-im\theta}\,d\theta\end{array}.$$

Let us consider the function $\tilde{h}=h\circ\cos$.
This function can be extended in a periodic way to $\mathbb{R}$
and it is continuous within its interval of periodicity. Its
derivative is piecewise continuous in this interval, so we are
under the assumption of the theorem \ref{th1} and it can be stated
that there exists a constant $B>0$ such that
$$|\widehat{\tilde{h}}[l]|\leq \frac{B}{l^2}, \hspace{0.5cm}\forall\,l\in\mathbb{N}.$$

Working in a similar way, we reach the following
relation
$$f_{nm}=\int\limits_{-1}^1\left[\int\limits_{-\pi}^\pi f\left(\frac{1+z}{1-z}
\,L,\theta\right)\e^{-im\theta}\,d\theta\right]
\,T_{n}(z)(1-z^2)^{-1/2}dz=h_{n}.$$

Using the proposition \ref{prop1}, we have that
$\displaystyle \widehat{\tilde{h}}[n]=\frac{1}{k}\,h_{n}$,
then
\begin{equation}
|f_{nm}|\leq\frac{B}{n^2}.\label{acot2}
\end{equation}

Finally, using (\ref{acot1}) and (\ref{acot2}), it is
obtained
$$|f_{nm}|^2\leq\frac{AB}{|m|^{p+1}n^2}.$$
\end{demo}

\section{Sturm-Liouville equations}

\label{apendice2}\rm

\noindent The so called \textit{Sturm-Liouville} differential
equations fit the following expression
\begin{equation}
\left[r(x)\,y'\right]'+\left[q(x)+\lambda p(x)\right]\,y=0.
\end{equation}

\noindent The solutions $y(x)$ of this ordinary differential
equation are called \textit{eigenfunctions} while the valid values
$\lambda$ associated to them receive the name of
\textit{eigenvalues}. It can be proved that if the functions $p$,
$q$ and $r$ are real and continuous over an interval
$\left[a,b\right]$ being $p(x)>0$ (or $p(x)<0$),$\forall\,
x\in\,[a,b]$, then all the eigenvalues $lambda$ are real and the
eigenfunctions $y$ are orthogonal in $[a,b]$ with the weight
function $p$.\\

\noindent It is easy to see that the rational Chebyshev functions
constitutes a family of eigenfunctions of a Sturm-Liouville
equations, specifically the $\left\{TL_n(x)\right\}_n$ functions
form the solutions set of the following EDO
\begin{equation}
\left(\left(x+L\right)\sqrt{\frac{x}{L}}\,y'(x)\right)'+\frac{n^2}{x+L}\;
\sqrt{\frac{L}{x}}\;y(x)=0,
\end{equation}

\noindent displaying, therefore, all the properties of the
Sturm-Liouville eigenfunctions.\\

\section{Linear transformations}
\label{apendice3}

It is important to know how the coefficients behave under some linear transformations in the original galaxy image, such as rotations, dilation or contractions or an increasing of the flux. due to the linearity of the CHEF functions, calculation the translation of these transformations into CHEF domain is straightforward.
\begin{itemize}
\item \textit{Rotation.} Given $f^{rot}(r,\theta):=f(r,\theta+\rho)$ with $\rho\in{[-\pi,\pi]}$ any angle, it can be proved that
$$f_{nm}^{rot}=f_{nm}\e^{im\rho}.$$
That is, a rotation of $\rho$ radians in real space is equivalent to a rotation of $m\rho$ radians in CHEF space.
 \item \textit{Dilation or contraction.} Given  $\alpha\in\mathbb{R}$ and defining $f^{dil}(r,\theta):=f(\alpha r,\theta)$, it yields
 $$f_{nm;L}^{dil}=f_{nm;\alpha L}.$$
What indicates that a dilation or contraction by a factor $\alpha$ in real space is equivalent to a dilation or contraction by the same factor in the parameter scale $L$ in CHEF space.
\item \textit{Flux increasing.} Given $k\in\mathbb{R}$ and $f^{inc}(r,\theta):=k\cdot f(r,\theta)$, it is obtained
$$f_{nm}^{inc}=k\cdot f_{nm}.$$
This means that an increasing in the flux by a factor $k$ is equivalent to an increasing in the CHEF coefficients by the same factor.
\end{itemize}

\end{document}